\documentclass[a4paper,11pt]{article}
\pdfoutput=1 

\usepackage{jcappub} 

\usepackage[T1]{fontenc} 

\title{Semi-dynamical perturbations of unified dark energy}

\author[a]{Lucas~Lombriser,}
\author[a]{Andy~Taylor}

\affiliation[a]{Institute for Astronomy, University of Edinburgh, Royal Observatory, Blackford Hill, Edinburgh, EH9~3HJ, U.K.}

\emailAdd{llo@roe.ac.uk}
\emailAdd{ant@roe.ac.uk}

\newcommand{\bq}{\begin{equation}}
\newcommand{\eq}{\end{equation}}
\newcommand{\bqa}{\begin{eqnarray}}
\newcommand{\eqa}{\end{eqnarray}}

\newcommand{\rmd}{{\rm d}}
\newcommand{\hMpc}{h^{-1}{\rm Mpc}}
\newcommand{\Om}{\Omega_{\rm m}}
\newcommand{\rhom}{\rho_{\rm m}}
\newcommand{\Deltam}{\Delta_{\rm m}}
\newcommand{\Vm}{V_{\rm m}}
\newcommand{\fPhi}{f_{\Phi}}

\newcommand{\fzeta}{f_{\zeta}}
\newcommand{\fDelta}{f_{\Delta}}
\newcommand{\Fcal}{\mathcal{F}}

\newcommand{\Sm}{S_{\rm m}}
\newcommand{\delg}{\delta g^{00}}
\newcommand{\Mbar}{\bar{M}}
\newcommand{\Mhat}{\hat{M}}

\newcommand{\aK}{\alpha_{\rm K}}
\newcommand{\aM}{\alpha_{\rm M}}
\newcommand{\aB}{\alpha_{\rm B}}
\newcommand{\aT}{\alpha_{\rm T}}
\newcommand{\aH}{\alpha_{\rm H}}
\newcommand{\cs}{c_{\rm s}}

\newcommand{\bontwt}{\tilde{\beta}_{12}}
\newcommand{\btht}{\tilde{\beta}_{3}}
\newcommand{\bont}{\tilde{\beta}_{1}}
\newcommand{\bonfot}{\tilde{\beta}_{14}}
\newcommand{\bonfit}{\tilde{\beta}_{15}}
\newcommand{\bonsit}{\tilde{\beta}_{16}}
\newcommand{\bset}{\tilde{\beta}_{7}}
\newcommand{\bonsieit}{\tilde{\beta}_{168}}
\newcommand{\bsenit}{\tilde{\beta}_{79}}
\newcommand{\gon}{\gamma_1}
\newcommand{\gni}{\gamma_9}
\newcommand{\gnit}{\tilde{\gamma}_9}

\newcommand{\en}{\epsilon_{\rm n}}
\newcommand{\eon}{\epsilon_1}
\newcommand{\etw}{\epsilon_2}
\newcommand{\ethr}{\epsilon_3}
\newcommand{\efo}{\epsilon_4}
\newcommand{\efi}{\epsilon_5}
\newcommand{\esi}{\epsilon_6}
\newcommand{\ese}{\epsilon_7}
\newcommand{\eonsiei}{\epsilon_{168}}
\newcommand{\eseni}{\epsilon_{79}}
\newcommand{\eBon}{\epsilon_{{\rm B}1}}
\newcommand{\eBthr}{\epsilon_{{\rm B}3}}
\newcommand{\eBfo}{\epsilon_{{\rm B}4}}

\abstract{
 Linear cosmological perturbations of a large class of modified gravity and dark energy models can be unified in the effective field theory of cosmic acceleration, encompassing Horndeski scalar-tensor theories and beyond.
 The fully available model space inherent to this formalism cannot be constrained by measurements in the quasistatic small-scale regime alone.
 To facilitate the analysis of modifications from the concordance model beyond this limit, we introduce a semi-dynamical treatment extrapolated from the evolution of perturbations at a pivot scale of choice.
 At small scales, and for Horndeski theories, the resulting modifications recover a quasistatic approximation but account for corrections to it near the Hubble scale.
 For models beyond Horndeski gravity, we find that the velocity field and time derivative of the spatial metric potential can generally not be neglected, even in the small-scale limit.
 We test the semi-dynamical approximation against the linear perturbations of a range of dark energy and modified gravity models, finding good agreement between the two.
}

\begin{document}
\maketitle
\flushbottom


\section{Introduction} \label{sec:intro}

Identifying the nature of the late-time accelerated expansion of our Universe is a prime endeavour of cosmology.
Instead of a vacuum energy in the form of an inexplicably small cosmological constant, cosmic acceleration may be driven by the contribution of a dark energy or a modification of gravity.
Conceptually, gravitational physics is not understood in the ultraviolet and a more fundamental theory of gravity may give rise to a remnant in the infrared that could cause the effect.
The simplest potential contribution is a single low-energy effective scalar degree of freedom.
A plethora of modified gravity models have been proposed based on the prospects of such a field~\cite{clifton:11,joyce:14,koyama:15} and a more systematic approach to explore their cosmological implications has become a necessity.
In the past few years, extensive efforts have been conducted to develop a generalised formalism for the resulting modifications in the formation of structure~\cite{uzan:06,caldwell:07,zhang:07,amendola:07,hu:07b,amin:07,bertschinger:08,daniel:10,pogosian:10,bean:10,park:10,hojjati:11,battye:12,baker:12,gubitosi:12,bloomfield:12,lombriser:13a,tsujikawa:14,bellini:14,gleyzes:14b}.
The effective field theory (EFT) of cosmic acceleration~\cite{creminelli:08,park:10,gubitosi:12,bloomfield:12,gleyzes:13,piazza:13a,hu:13b,tsujikawa:14,bellini:14,lombriser:14b,gleyzes:14b}, or unified dark energy, provides such a framework and describes the evolution of the spatially homogeneous and isotropic background of our Universe and the perturbations around it for a large class of modified gravity and dark energy scenarios.

Solving the coupled system of linearly perturbed modified Einstein equations for general dark energy and modified gravity models can be a time-consuming numerical challenge and may only offer little insight into the generic behaviour of the wide range of possible modifications.
Contrary to the $\Lambda$ Cold Dark Matter ($\Lambda$CDM) concordance model, where the evolution of the gravitational potential is scale independent, for modified models the field and conservation equations need to be re-evaluated at each scale of interest.
To circumvent the consequent increase in computational demand, and to enable analytic calculations, frequently a quasistatic approximation is employed, which neglects time derivatives compared to spatial derivatives and usually emphasises density perturbations compared to large-scale velocity currents in the field equations.
This can provide direct information about the relations between the fluctuations without requiring the integration of the field and conservation equations.
However, the contributions of time derivatives of the metric potentials and the velocities can become important near the Hubble scale, or near the sound horizon for models with subluminal sound speed in the dark energy component.
These effects can limit tests of gravity and dark energy employing the quasistatic approximation and including observations at very large scales~\cite{lombriser:13a,duniya:13,sawicki:15}.
Importantly, one of the free time-dependent functions characterising the EFT model space does not enter the quasistatic computations~\cite{lombriser:14b} and, hence, cannot be constrained by measurements in the quasistatic regime.
An analysis beyond the quasistatic limit is also required to lift degeneracies between the standard concordance cosmology and other subsets of the EFT model space~\cite{lombriser:14b}.
Moreover, the cosmological background and the linear leading-order modifications at small scales, characterised by a deviation in the Poisson equation and a gravitational slip between the metric potentials, only allow a measurement of three independent functions of time.
In principle, a measurement of the linear propagation speed and friction of gravitational waves could be used to further restrict the available model space~\cite{saltas:14,amendola:14,raveri:14b,pettorino:14} but the required observations associated with a late-time cosmological modification are currently lacking.
However, since the EFT formalism contains more freedom than this, in order to fully constrain the available model space, next-to-leading-order contributions to these modifications need to be measured.
This implies the consideration of the velocity field and the time derivatives of the fluctuations.
Hence, a more general, yet efficient approach beyond the quasistatic approximation that also allows for direct analytical operations to gain insights on the rich phenomenology of dark energy and modified gravity models is highly desirable.

In this paper, we adopt the unified dark energy formalism of Ref.~\cite{bellini:14,gleyzes:14b}, encompassing Horndeski scalar-tensor theories~\cite{horndeski:74,deffayet:11} and models beyond it~\cite{gleyzes:14a}, and explore the linear cosmological perturbations that it describes.
We extend previous work by deriving the energy-momentum conservation equations and a set of combined modified Einstein equations where the scalar field fluctuations are eliminated and which incorporate beyond-Horndeski models.
We then introduce a semi-dynamical treatment extrapolated from the evolution of the perturbations at a chosen scale, the pivot scale, to facilitate the analysis of modifications from the concordance model and its application to observational tests of gravity and dark energy.
We describe the resulting effective deviation in the Poisson equation, the gravitational slip, and an approximated growth rate of matter density fluctuations.
This also includes a discussion on the choice of pivot scale and its impact on the computation of these relations.
Finally, we numerically test the semi-dynamical approximation against the exact linear perturbations for a range of dark energy and modified gravity models that are embedded in the Horndeski and beyond-Horndeski action.
These include models with a kinetic contribution of the scalar field, a running Planck mass, kinetic braiding, deviations between the speed of gravitational waves and the speed of light, as well as beyond-Horndeski terms.
We also compare our results against a quasistatic approximation.

We begin by briefly reviewing the unified dark energy formalism in Sec.~\ref{sec:ude}. 
In Sec.~\ref{sec:semidynamics}, we then introduce the semi-dynamical approximation for linear cosmological perturbations of unified dark energy, determining the time and scale dependence of the effective modifications. 
We test the semi-dynamical approximation against numerical solutions of the exact modified Einstein and energy-momentum conservation equations in Sec.~\ref{sec:examples} and discuss our results in Sec.~\ref{sec:conclusion}.
The modified linearly perturbed Einstein equations and coefficients of the effective modifications are given in the appendices for reference.


\section{Unified dark energy} \label{sec:ude}

The effective field theory of cosmic acceleration~\cite{gubitosi:12,bloomfield:12} provides a unified framework for describing the expansion of a Friedmann-Lema\^itre-Robertson-Walker (FLRW) background and perturbations around it in a generalised theory of gravity constructed from supplementing the tensor field with a single scalar field and with the matter sector obeying the weak equivalence principle.
Adopting the unitary gauge, where the time coordinate is chosen in order to absorb the scalar field perturbation in the metric $g_{\mu\nu}$, the unified treatment reduces to finding a generalised action composed of geometric operators that are invariant under time-dependent spatial diffeomorphisms with free time-dependent coefficients.
The EFT action, up to quadratic order, becomes~\cite{gubitosi:12,bloomfield:12}
\bqa
 S & = & \frac{1}{2\kappa^2} \int d^4x\sqrt{-g} \left[ \phantom{\frac{}{}} \Omega(t) R - 2\Lambda(t) - \Gamma(t) \delg + M_2^4(t) (\delg)^2 - \Mbar_1^3(t) \delg \delta K^{\mu}_{\ \mu} \right. \nonumber\\  & & \left. - \Mbar_2^2(t)(\delta K^{\mu}_{\ \mu})^2 - \Mbar_3^2(t) \delta K^{\mu}_{\ \nu} \delta K^{\nu}_{\ \mu} + \Mhat^2(t) \delg \delta R^{(3)} + m_2^2(t)(g^{\mu\nu}+n^{\mu}n^{\nu})\partial_{\mu} g^{00} \partial_{\nu} g^{00} \right] \nonumber\\
& & + \Sm\left[\psi_{\rm m};g_{\mu\nu}\right], \label{eq:eftaction}
\eqa
where $\kappa^2\equiv8\pi\,G$ with bare gravitational constant $G$ and the speed of light in vacuum is set to unity.
Furthermore, $R$ and $R^{(3)}$ denote the four-dimensional and spatial Ricci scalar, respectively, $K_{\mu\nu}$ is the extrinsic curvature tensor, and $n^{\mu}$ describes the normal to surfaces of constant time, whereby $\delta$ denotes perturbations with respect to the background.
We have adopted the notation of Ref.~\cite{lombriser:14b} for the EFT coefficients.
Moreover, we only considered the leading order contributions with respect to their mass dimension in Eq.~(\ref{eq:eftaction}) and neglected terms that introduce higher-order time derivatives in the Euler-Lagrange equations.

Given the quadratic order, Eq.~(\ref{eq:eftaction}) fully describes the background evolution and linear perturbations of an extensive class of scalar-tensor theories.
In particular it encompasses Horndeski theory~\cite{horndeski:74,deffayet:11}, which describes the most general local, Lorentz-covariant, and four-dimensional scalar-tensor gravity where the Euler-Lagrange equations are at most second-order in the derivatives of the scalar and tensor fields.
In Horndeski gravity, all coefficients, except for $m_2$, are used in Eq.~(\ref{eq:eftaction}) with the restriction that $\Mbar_2^2=-\Mbar_3^2=2\Mhat^2$.
Healthy theories beyond the Horndeski action~\cite{gleyzes:14a}, introducing third-order derivatives in the modified Einstein equations but with a constraint equation ensuring a second-order equation for the propagating scalar degree of freedom, are also embedded in Eq.~(\ref{eq:eftaction}), allowing $\Mbar_2^2\neq2\Mhat^2$.
Finally, a nonzero $m_2$ is, for instance, introduced in Ho\v{r}ava-Lifshitz~\cite{horava:09} gravity with the violation of Lorentz covariance.
The concordance model phenomenology is recovered when $\Omega=1$, $\Lambda$ is a constant, and the remaining coefficients vanish, but it can be degenerate with other choices of the EFT functions on quasistatic scales~\cite{lombriser:14b}.
In general, the EFT action, Eq.~(\ref{eq:eftaction}), introduces nine time-dependent coefficients to which the statistically spatially homogeneous and isotropic background metric adds the scale factor $a(t)$, or the Hubble parameter $H(t)$.
The Friedmann equations introduce two constraints such that the formalism is composed of eight free time-dependent functions, given the matter content and spatial curvature $k_0$~\cite{gubitosi:12,bloomfield:12,lombriser:14b}.
Importantly, the EFT function $M_2^4(t)$ is not contributing in the quasistatic limit~\cite{lombriser:14b}.

In the following, we study the linear scalar perturbations around the FLRW background metric in unified dark energy, restricting to a spatially flat ($k_0=0$) universe, otherwise containing only pressureless dust $p_{\rm m}=0$.
We furthermore restrict to models where $\Mbar_2^2=-\Mbar_3^2$ and $m_2=0$ in the EFT action, Eq.~(\ref{eq:eftaction}).
This encompasses Horndeski gravity, where additionally $2\Mhat^2=\Mbar_2^2$, and beyond-Horndeski models with the corresponding effective theory described by five and six free functions of time, respectively.
We adopt the Newtonian gauge with the FLRW line element described by
\bq
 \rmd s^2 = -(1+2\Psi)\rmd t^2 + a^2(t)(1+2\Phi)\rmd{\bf x}^2 \label{eq:FLRW}
\eq
and matter fluctuations are considered in the total matter gauge with velocities $\Vm$ and density fluctuations $\Deltam$ around the background matter density $\rhom$.
However, instead of working with the effective field theory coefficients of Eq.~(\ref{eq:eftaction}), we choose to adopt a description that characterises the free time-dependent functions of the unified dark energy with more direct physical, and observational, implications.
Such a framework was developed in Ref.~\cite{bellini:14} describing the background and linear perturbation theory equivalent to Horndeski gravity.
We adopt the slightly different notation of Ref.~\cite{gleyzes:14b}, thereby also employing their extension that incorporates beyond-Horndeski models.
The formalism separates out the expansion history $H$ as the free function determining the cosmological background, which relates to the EFT functions $\Omega$, $\Gamma$, and $\Lambda$ via the Friedmann equations.
Linear perturbations around the background are then characterised by five additional free functions of time $\alpha_i$, each implying a different physical effect in the properties of the unified dark energy.
\begin{itemize}
 \item {\bf Kineticity $\aK$:} The contribution of a kinetic energy of the scalar field can give rise to a clustering of the dark energy component at very large scales.
  It relates to Eq.~(\ref{eq:eftaction}) as
  \bq
    \aK = \frac{\Gamma+4M_2^4}{H^2 (\Omega+\Mbar_2^2)}.
  \eq
 \item {\bf Planck mass evolution rate $\aM$:} The evolution of the gravitational coupling gives rise to a gravitational slip between the metric potentials $\Psi$ and $\Phi$.
  This attributes an effective anisotropic stress to the dark energy component.
  In terms of the EFT functions of Eq.~(\ref{eq:eftaction}), the parameter is expressed as
  \bq
   \aM = \frac{\Omega' + (\Mbar_2^2)'}{\Omega + \Mbar_2^2},
  \eq
  where primes denote derivatives with respect to $\ln a$ here and throughout the paper.
  The Planck mass is $M^2 = \kappa^{-2}(\Omega + \Mbar_2^2)$ with $\aM = \rmd \ln M^2/\rmd \ln a$ and relates to the background expansion via $H = \aM^{-1} \rmd\ln M^2/\rmd t$.
 
 \item {\bf Braiding $\aB$:} The interaction of the scalar field and the metric through braiding, or mixing, of the kinetic contributions of these fields causes the dark energy component to cluster at small scales.
  The relation to the EFT framework is given by
  \bq
   \aB = \frac{H\Omega' + \Mbar_1^3}{2H (\Omega+\Mbar_2^2)}.
  \eq
 \item {\bf Tensor speed alteration $\aT$:} A deviation between the speed of gravitational waves and the speed of light contributes as an effective anisotropic stress and clustering of the dark energy component.
  The parameter relates to Eq.~(\ref{eq:eftaction}) via
  \bq
   \aT = -\frac{\Mbar_2^2}{\Omega+\Mbar_2^2}.
  \eq
 \item {\bf Beyond-Horndeski term $\aH$:} The introduction of third-order derivatives in the Einstein equations, however, with a second-order equation for the propagating scalar degree of freedom, lies outside of the model space of Horndeski gravity~\cite{gleyzes:14a}.
  As will be discussed in Sec.~\ref{sec:semidynamics}, this causes velocity fields and the rate of evolution of the metric potentials to contribute to the anisotropic stress and clustering of the dark energy on small scales at leading order.
  The relation to the EFT formalism is provided by
  \bq
   \aH = \frac{2\Mhat^2-\Mbar_2^2}{\Omega+\Mbar_2^2}.
  \eq
\end{itemize}
Note that the functions $\alpha_i$ can also directly be defined from the quadratic expansion of the action composed in terms of geometric quantities invariant under spatial diffeomorphisms in the $3+1$ Arnowitt-Deser-Misner (ADM) spacetime decomposition with uniform scalar field hypersurfaces of constant time~\cite{gleyzes:14b}.

In order to describe the cosmological perturbations implied by Eqs.~(\ref{eq:eftaction}) and (\ref{eq:FLRW}), the time diffeomorphism $t\rightarrow t+\pi(t,{\bf x})$ with fluctuation of the scalar degree of freedom $\pi$ is applied to the action to restore its full four-dimensional covariance.
We work in Fourier space and simplify notation by referring to perturbative quantities through their Fourier amplitudes of the plane waves with comoving wavenumber $k$, where in linear theory the phases factor out in the field equations.
We refer to Ref.~\cite{gleyzes:14b} or Appendix~\ref{sec:einsteineqs} for the resulting perturbed modified Einstein equations.
We follow Ref.~\cite{bellini:14} to combine the field equations in two modified Einstein equations where contributions of $\pi$, its derivatives, and $\Psi'$ are eliminated.
This implies using $\pi''$ determined by the scalar field equation, Eq.~(\ref{eq:scalarfieldeq}), in the time-space component Eq.~(\ref{eq:timespace}) to get $\pi'$,
which then is used in the traceless space-space component Eq.~(\ref{eq:tracelessspacespace}) to obtain $\pi$.
The time-time component Eq.~(\ref{eq:timetime}) then determines $\Psi$, while $\Psi'$ can be obtained from the time derivative of the traceless space-space component.
Finally, we combine these results in the trace of the space-space component Eq.~(\ref{eq:tracespacespace}) to get the first combined modified Einstein equation,
\bqa
 \Phi'' + A_1 \Phi' + A_2 k_H^2 \Phi = \rhom \left(A_3 \Deltam + A_4 \frac{\Vm}{k_H}\right), \label{eq:einstein1}
\eqa
where the coefficients are given by
\bqa
 A_1 & = & \frac{H'}{H} + \frac{\bontwt + \btht (\aB - \aH)^2 k_H^2}{\bont + (\aB - \aH)^2 k_H^2}, \\
 A_2 & = & \frac{\bonfot k_H^{-2}+\bonfit+(\aB-\aH)^2 \cs^2 k_H^2}{\bont+(\aB-\aH)^2 k_H^2}, \\
 A_3 & = & \frac{1}{2H^2 M^2} \frac{\bonsit+\bset k_H^2}{\bont+(\aB-\aH)^2 k_H^2}, \\
 A_4 & = & -\frac{1}{2H^2 M^2} \frac{\bonsieit+\bsenit (\aB-\aH)^2 k_H^2+\aH \en k_H^4}{\bont+(\aB-\aH)^2 k_H^2}
\eqa
with $k_H\equiv k/(aH)$ and
\bqa
 \tilde{\beta}_i & = & \beta_i + \aH \epsilon_i, \label{eq:betatilde1} \\
 \tilde{\beta}_{1j} & = & \beta_1 \beta_j + \aH \epsilon_j, \\
 \bonsieit & = & \beta_1 \left(3 \beta_6 + \beta_8\right) + \aH \eonsiei, \\
 \bsenit & = & 3 \beta_7 + \beta_9 + \aH \eseni \label{eq:betatilde2}
\eqa
for $i=1,3,7$ and $j=2,4,5,6$.
The parameters $\beta_i$ are independent of $\aH$.
They are defined in the appendices of Refs.~\cite{bellini:14,gleyzes:14b} and, hence, shall not be given here again.
In Eqs.~(\ref{eq:betatilde1}) to (\ref{eq:betatilde2}), we extend this parametrisation to theories beyond Horndeski gravity ($\aH\neq0$), introducing the parameters $\epsilon_i$ defined in Appendix~\ref{sec:coefficients}.
Finally, we have also made use of the sound speed of the dark energy component~\cite{bellini:14,gleyzes:14b}
\bq
 \cs^2 = -\frac{2 (1 + \aB)^2}{\alpha} \left[1 + \aT - \frac{1 + \aH}{1 + \aB} \left(1 + \aM - \frac{H'}{H}\right) - \left(\frac{1 + \aH}{1 + \aB}\right)'\right] - \frac{(1 + \aH)^2}{\alpha} \frac{\rhom}{H^2 M^2}, \label{eq:soundspeed}
\eq
where $\alpha\equiv6\aB^2+\aK$.

A second field equation follows from using $\pi'$ determined from the time-time component Eq.~(\ref{eq:timetime}) in the time-space component Eq.~(\ref{eq:timespace}) to get $\pi$.
From the traceless space-space component Eq.~(\ref{eq:tracelessspacespace}), it then follows that
\bq
 \Phi' + k_H^2 ( B_1 \Phi + B_2 \Psi ) = \rhom \left(B_3 \Deltam + B_4 \frac{\Vm}{k_H}\right), \label{eq:einstein2}
\eq
where the coefficients are given by
\bqa
 B_1 & = & \frac{(1+\aT) \left(\gon k_H^{-2}+\aB^2\right)+\frac{2\aB}{\alpha}\gni + \aH \eBon}{\gnit + \aH (\aB-\aH)k_H^2}, \\
 B_2 & = & \frac{\bont k_H^{-2} + (\aB-\aH)^2}{\gnit+\aH (\aB-\aH) k_H^2}, \\
 B_3 & = & \frac{1}{H^2 M^2}\frac{\frac{\aB}{\alpha}\gni + \aH\eBthr}{\gnit+\aH (\aB-\aH) k_H^2}, \\
 B_4 & = & -\frac{1}{2H^2 M^2}\frac{\frac{\aK}{\alpha}\gni+\aH \eBfo+\aH(\aB-\aH) k_H^2}{\gnit+\aH (\aB-\aH) k_H^2}
\eqa
with $\gnit = \gni - \aH \eon$ and $\epsilon_{{\rm B}i}$ defined in Appendix~\ref{sec:coefficients}.
The parameters $\gamma_i$ are independent of $\aH$.
They can be found in Refs.~\cite{bellini:14,gleyzes:14b} and are not given here again.

Finally, in addition to Eqs.~(\ref{eq:einstein1}) and (\ref{eq:einstein2}), we derive the energy-momentum conservation equations from the modified Einstein equations in Appendix~\ref{sec:einsteineqs}, which will close the system of differential equations.
Using the time derivatives of the time-space and trace of the space-space components to eliminate $\Phi''$ with $\pi'$ from the time-space component, we find the usual momentum conservation equation
\bq
 \Vm' + \Vm = k_H\Psi \label{eq:momcon}.
\eq
Using all of the modified Einstein equations and the scalar field equation together with the time derivatives of the time-time, time-space, and traceless space-space components to solve for $\Delta'$, $\Delta$, $\pi$ and $\Psi$ and their time derivatives, as well as $\Phi''$, we derive the standard energy conservation equation
\bqa
 \Deltam' = -k_H\Vm - 3\zeta', \label{eq:encon}
\eqa
where we have used the comoving curvature $\zeta=\Phi-\Vm/k_H$.


\section{Semi-dynamical linear cosmological perturbations} \label{sec:semidynamics}

As emphasised in Secs.~\ref{sec:intro} and \ref{sec:ude}, performing a quasistatic approximation of the modified Einstein and energy-momentum conservation equations, Eqs.~(\ref{eq:einstein1}), (\ref{eq:einstein2}), (\ref{eq:momcon}), and (\ref{eq:encon}), is not sufficient for conducting a more efficient but comprehensive test of gravity.
The approach is limited by its incomplete description at very large scales and cannot constrain the fully available EFT model space.
An analysis beyond the simple quasistatic approximation is therefore needed, implying the consideration of velocities and time derivatives of the metric fluctuations.
In order to account for the contribution of these terms in the field equations, we propose an extension of the quasistatic approximation that introduces corrections from $\Phi''$, $\Phi'$, and $\Vm/k_H$ in Eqs.~(\ref{eq:einstein1}) and (\ref{eq:einstein2}), determined at a pivot scale $k_*$ of choice.
More specifically, we solve the coupled system of modified Einstein and energy-momentum conservation equations at $k_*$ and then use this result to extrapolate relations between the perturbations to other scales by performing the replacements
\bqa
 \Phi' & \rightarrow & \fPhi\Phi, \label{eq:fPhi} \\
 \zeta'& \rightarrow & \fzeta\Phi \label{eq:fzeta}
\eqa
in Eqs.~(\ref{eq:einstein1}) and (\ref{eq:einstein2}), where $\fPhi\equiv\Phi'(k_*)/\Phi(k_*)$ and $\fzeta=\zeta'(k_*)/\Phi(k_*)$ characterise the rates of evolution of the spatial metric potential and the comoving curvature at the pivot scale, respectively.
Furthermore, from momentum conservation Eq.~(\ref{eq:momcon}), we obtain
\bqa
 \frac{\Vm}{k_H} & \rightarrow & \frac{H}{H'}\left[(\fPhi-\fzeta)\Phi-\Psi\right]. \label{eq:velocity}
\eqa
These replacements allow us to eliminate time derivatives of $\Phi$ and the velocity field in the modified Einstein equations.

Motivated by the energy-momentum conservation equations in Sec.~\ref{sec:ude} and the gauge transformation of the comoving curvature, which are not modified in unified dark energy, we estimate that relations between the fluctuations scale as
\bq
 \zeta\sim\Phi\sim\Psi\sim\frac{\Vm}{k_H}\sim\frac{\Deltam}{k_H^2}. \label{eq:scaling}
\eq
Hence, the field equations and the coefficients $A_i$ and $B_i$ suggest that the replacements proposed in Eqs.~(\ref{eq:fPhi}) through (\ref{eq:velocity}) will provide a good approximation for Horndeski scalar-tensor theories at small scales as the velocity field and the time derivatives of the metric potentials are subdominant and do not contribute in the limit of $k\rightarrow\infty$.
However, if we allow $\aH\neq0$, a dependency on $\Phi'$ at $k\rightarrow\infty$ enters through Eq.~(\ref{eq:einstein2}) and a dependency on $\Vm/k_H$ through Eqs.~(\ref{eq:einstein1}) and (\ref{eq:einstein2}), suggesting that a quasistatic approximation with $\fPhi=\fzeta=0$ becomes invalid (cf.~\cite{defelice:15}).
Hence, for beyond-Horndeski models nonzero contributions of $\fPhi$ and $\fzeta$ should be considered at leading order.
Note that here and throughout the paper, $k\rightarrow\infty$ refers to a formal limit taken in linear theory.
In concordance cosmology linear predictions are typically applicable to scales of $k\lesssim0.1~\hMpc$.
This limit can, however, be affected by modifications of gravity~\cite{lombriser:13c} and one should check against nonlinear corrections when using linear predictions to constrain the EFT model space with cosmological observations.

Alternatively to the replacement in Eq.~(\ref{eq:fzeta}), one could define $f_{\Delta}\Deltam\equiv\Deltam'$ and use Eq.~(\ref{eq:encon}) to eliminate velocities in the field equations at small scales and derive a first-order inhomogeneous nonlinear differential equation for $f_{\Delta}$ from the energy-momentum conservation equations and Eqs.~(\ref{eq:einstein1}) and (\ref{eq:einstein2}).
This would provide a good approximation at small scales but not facilitate computations at large scales.
For an approximation at super-Hubble scales ($k\rightarrow0$), we could also use $\fzeta$ as defined in Ref.~\cite{hu:07b} to replace velocities, i.e., $\lim_{k_H\rightarrow0} \zeta' = \fzeta k_H \Vm/3$.
However, this comes at the cost of introducing an extra scale dependency in Eq.~(\ref{eq:velocity}) that consequently adds further coefficients in the expansion of Eqs.~(\ref{eq:einstein1}) and (\ref{eq:einstein2}) in powers of $k$ and, hence, in the semi-dynamical description of effective unified dark energy modifications discussed in the following.


\subsection{Effective modifications} \label{sec:mod}

Performing the replacements Eqs.~(\ref{eq:fPhi}) through (\ref{eq:velocity}) in the modified Einstein equations of the Fourier amplitudes of linear cosmological fluctuations, Eqs.~(\ref{eq:einstein1}) and (\ref{eq:einstein2}), we obtain
\bqa
 \left[\fPhi' + \fPhi^2 + \left(A_1 - A_4 \frac{H}{H'}\rhom \right) \fPhi + A_4\frac{H}{H'}\rhom\fzeta + A_2 k_H^2\right] \Phi + A_4 \frac{H}{H'}\rhom \Psi & = &  A_3\rhom\Deltam, \label{eq:semidyneinst1} \\
 \left[\left(1-B_4\frac{H}{H'}\rhom\right)\fPhi + B_4\frac{H}{H'}\rhom\fzeta + B_1 k_H^2\right] \Phi + \left(B_4\frac{H}{H'}\rhom+B_2 k_H^2\right) \Psi & = & B_3\rhom\Deltam. \label{eq:semidyneinst2}
\eqa
Having cast the field equations in this form, we combine them and parametrise the gravitational properties of unified dark energy as an effective modification of the Poisson equation $\mu$ and by the introduction of a gravitational slip $\gamma$,
\bqa
 \mu(a,k) & \equiv & -\frac{2H^2k_H^2\Psi}{\kappa^2\rhom\Deltam} = \frac{1}{\kappa^2M^2} \frac{\mu_{+2}k_H^2 + \mu_{+4}k_H^4 + \mu_{+6} k_H^6}{\mu_{-0} + \mu_{-2}k_H^2 + \mu_{-4}k_H^4 + \mu_{-6}k_H^6}, \label{eq:mu} \\
 \gamma(a,k) & \equiv & -\frac{\Phi}{\Psi} = \frac{\gamma_{+0} + \gamma_{+2} k_H^2 + \gamma_{+4} k_H^4}{\mu_{+2} + \mu_{+4} k_H^2 + \mu_{+6} k_H^4}, \label{eq:gamma}
\eqa
respectively, where $\mu_{\pm 2n}$ and $\gamma_{+2n}$ ($n=0,1,2,3$) are functions of time only and presented in Appendix~\ref{sec:coefficients}.
In combination with energy-momentum conservation, Eqs.~(\ref{eq:momcon}) and (\ref{eq:encon}), and given $\fPhi$ and $\fzeta$, these two relations completely describe the linear fluctuations of unified dark energy in the semi-dynamical approximation and constitute the core of the approach from which further relations can be derived.

In the limit of $k\rightarrow\infty$, Eqs.~(\ref{eq:mu}) and (\ref{eq:gamma}) simplify to
\bqa
 \mu_{\infty} & = & \frac{\mu_{+6}}{\mu_{-6}} = \frac{1}{\kappa^2 M^2} \frac{\mu_{\infty}^+ + \aH \left( \fPhi\mu_{\Phi,\infty}^+ + \fzeta\mu_{\zeta,\infty}^+ \right)}{\mu_{\infty}^- + \aH \left( \fPhi\mu_{\Phi,\infty}^- + \fzeta\mu_{\zeta,\infty}^-\right)}, \\
 \gamma_{\infty} & = & \frac{\gamma_{+4}}{\mu_{+6}} = \frac{\gamma_{\infty}^+}{\mu_{\infty}^+ + \aH \left( \fPhi\mu_{\Phi,\infty}^+ + \fzeta\mu_{\zeta,\infty}^+ \right)},
\eqa
where the time-dependent coefficients are given by
\bqa
 \mu_{\infty}^+ & = & (\aB - \aH)^2 \left\{\alpha \left[\aB^2 (1 + \aT) \bset + \aH \left(\bset \eBon - 2 \eBthr\cs^2\right)\right] \right. \nonumber\\
  & & \left. + 2 \aB (\bset - \cs^2) \gni \right\} \frac{H'}{H}, \label{eq:muinfp} \\
 \mu_{\Phi,\infty}^+ & = & \alpha (\aB - \aH)^3 \bset \frac{H'}{H} - 2 \en (\alpha \aH \eBthr + \aB \gni) \frac{\rhom}{2 H^2 M^2}, \label{eq:muPhiinfp} \\
 \mu_{\zeta,\infty}^+ & = & \en (\alpha \aH \eBthr + \aB \gni) \frac{\rhom}{H^2 M^2}, \label{eq:muzetainfp} \\
 \mu_{\infty}^- & = & \alpha (\aB - \aH)^4 \cs^2 \frac{H'}{H} + \aH \en \left\{\alpha \left[\aB^2 (1 + \aT) + \aH \eBon\right] + 2 \aB \gni\right\} \frac{\rhom}{2 H^2 M^2}, \label{eq:muinfm} \\
 \mu_{\Phi,\infty}^- & = & \alpha \aB (\aB - \aH) \en \frac{\rhom}{2 H^2 M^2}, \label{eq:muPhiinfm} \\
 \mu_{\zeta,\infty}^- & = & -\alpha \en (\aB - \aH)^2 \frac{\rhom}{2 H^2 M^2}, \label{eq:muzetainfm} \\
 \gamma_{\infty}^+ & = & \alpha (\aB - \aH)^4 \bset \frac{H'}{H} + \aH \en (\alpha \aH \eBthr + \aB \gni) \frac{\rhom}{H^2 M^2}. \label{eq:gammainfp}
\eqa
For Horndeski theories, where $\aH=0$, this further simplifies to (cf.~\cite{bellini:14,gleyzes:14b})
\bqa
 \mu_{\infty} & = & \frac{1}{\kappa^2M^2}\frac{2\left[\aB(1+\aT)-\aM+\aT\right]^2 + \alpha(1+\aT) c_{\rm s}^2}{\alpha \cs^2}, \label{eq:muinfH} \\
 \gamma_{\infty} & = & \frac{2\aB \left[\aB(1+\aT)-\aM+\aT\right] + \alpha c_{\rm s}^2}{2\left[\aB(1+\aT)-\aM+\aT\right]^2 + \alpha (1+\aT) \cs^2}. \label{eq:gammainfH}
\eqa
As expected, while for Horndeski theories, $\mu_{\infty}$ and $\gamma_{\infty}$ are independent of $\fPhi$ and $\fzeta$, this is no longer true if $\aH\neq0$, implying that the quasistatic approach $\fPhi=\fzeta=0$ is generally not a good approximation for beyond-Horndeski theories.
Hence, when $\aH\neq0$, even in the limit of $k\rightarrow\infty$, neither $\Phi'$ in Eq.~(\ref{eq:einstein2}) nor $\Vm/k_H$ in Eqs.~(\ref{eq:einstein1}) and (\ref{eq:einstein2}) should be neglected.

It is also worth noting that a combined measurement of the cosmological background as well as $\mu_{\infty}$ and $\gamma_{\infty}$ through the large-scale structure only provides constraints on $H$ and two $\alpha_i$ functions.
This generally leaves two $\alpha_i$ functions unconstrained in Horndeski gravity and an additional one in beyond-Horndeski models.
In principle, one could obtain additional constraints on $\aT$ and $\aM$ from a measurement of the linear propagation speed and friction of gravitational waves~\cite{saltas:14,pettorino:14} but the required observations attributed to a late-time cosmological effect are not yet available.
Also note the independence of Eqs.~(\ref{eq:muinfH}) and (\ref{eq:gammainfH}) on $\aK$.
Hence, to fully constrain the available model space in the EFT formalism another two to three measurements are required, for instance of $\mu$ and $\gamma$ at additional, larger scales, where, however, $\Vm/k_H$ and time derivatives of $\Phi$ may not be neglected, and for which we envisage the use of the semi-dynamical expressions Eqs.~(\ref{eq:mu}) and (\ref{eq:gamma}).
Importantly, a quasistatic consideration of these additional scales instead suffers from degeneracies between specific relations in the EFT coefficients and concordance cosmology and contains no information on $M_2^4(t)$~\cite{lombriser:14b}.


\subsection{Growth of structure} \label{sec:growth}

In the semi-dynamical approach the logarithmic growth rate of matter density fluctuations $\fDelta=\Deltam'/\Deltam$ can in principle directly be determined from energy conservation and the evaluation of $\fPhi$ and $\fzeta$ at the pivot scale $k_*$.
However, in the determination of $\fDelta$, deviations from the pivot-scale evolution become important since velocities contribute at first order to $\Deltam'$ through energy conservation.
More specifically, from Eq.~(\ref{eq:encon}) and the replacements in Eqs.~(\ref{eq:fzeta}) and (\ref{eq:velocity}), we find
\bq
 \fDelta = -\mu \gamma \left[ \frac{H}{H'}\left(\fPhi-\fzeta + \frac{1}{\gamma} \right) + \frac{3\fzeta}{k_H^2} \right] \frac{\kappa^2\rhom}{2H^2}, \label{eq:fDelta}
\eq
where the modification of the Poisson equation $\mu$ and the gravitational slip $\gamma$ are given in Eqs.~(\ref{eq:mu}) and (\ref{eq:gamma}), respectively.
In the concordance model this reduces to the precise relation $\fDelta=1+\fPhi$.
The accuracy of Eq.~(\ref{eq:fDelta}) can, however, suffer for models introducing a strong scale dependency in $\fPhi$ and $\fzeta$, rendering a pivot-scale evaluation insufficient.

The logarithmic growth rate $\fDelta$ relates velocities to matter density fluctuations via energy conservation in Eq.~(\ref{eq:encon}).
The proportionality between the two becomes exact when $\zeta'\rightarrow0$.
To improve the accuracy in the recovery of this relation, instead of using energy conservation, we can adopt the parametrisation $\Fcal\Deltam=-k_H\Vm$~\cite{lombriser:13a}.
This allows us to formulate an evolution equation for $\Fcal$ using momentum conservation in Eq.~(\ref{eq:momcon}),
\bq
 \Fcal' + \Fcal^2 + \left(2+\frac{H'}{H} - \mu\gamma \frac{3\fzeta}{k_H^2}\frac{\kappa^2\rhom}{2H^2} \right)\Fcal - \mu \frac{\kappa^2\rhom}{2H^2} = 0, \label{eq:Fcal}
\eq
where we have used that $\Fcal$ and $\fDelta$ relate as
\bq
 \Fcal = \fDelta + \mu\gamma \frac{3\fzeta}{k_H^2}\frac{\kappa^2\rhom}{2H^2}.
\eq
Note that $\Fcal$ and $\fDelta$ agree in the limit of $k\rightarrow\infty$ and Eq.~(\ref{eq:Fcal}) reduces to the more familiar quasistatic differential equation for $\fDelta$.
Unlike that equation, however, Eq.~(\ref{eq:Fcal}) applies to all scales and is not limited to quasistatic scales only.

The parametrisation $\Fcal$ can be motivated by its observational implications, describing the relation of redshift-space distortions to the underlying matter density fluctuations, as these measure velocities rather than growth of structure $\fDelta$ directly.
One can then furthermore directly determine from $\Fcal$ the modifications~\cite{lombriser:13a} of the relativistic effects contributing to galaxy clustering~\cite{yoo:09,yoo:10,bonvin:11,challinor:11}, where a non-quasistatic treatment of perturbations can become essential.
These deviations can be large for models like Dvali-Gabadadze-Porrati braneworld gravity~\cite{dvali:00}, where $\gamma$ can be of order unity at the largest observable scales~\cite{hu:07b,lombriser:09}, or for linearly shielded modified gravity models~\cite{lombriser:14b}, which recover $\Lambda$CDM on quasistatic scales.
If measured, these effects may provide new insights to gravitational physics and dark energy near the Hubble scale.
Finally, Eq.~(\ref{eq:Fcal}) may also serve to formulate a growth-index parametrisation of matter density fluctuations $\tilde{\gamma}$~\cite{peebles:80,linder:05}, or rather of the velocity-to-density ratio, through $\Fcal=\Om(a)^{\tilde{\gamma}}$, where $\Om(a)\equiv\kappa^2\rhom/(3H^2)$.
This is a convenient parametrisation for observational tests of modified gravity and dark energy (e.g.,~\cite{samushia:12}), and in this form can be made consistent for applications beyond quasistatic scales (cf.~\cite{lombriser:11a}).
For small modifications of the concordance model, Eq.~(\ref{eq:Fcal}) may also be linearised in the deviations to simplify computations~\cite{baker:13}.


\subsection{Super-Hubble evolution} \label{sec:superH}

While the description of the modified structure is simplified in the limit of small scales, where as discussed in Secs.~\ref{sec:mod} and \ref{sec:growth} the approximation $\fPhi=\fzeta=0$ becomes accurate for Horndeski theories to leading order, a similar approximation can be made at super-Hubble scales.
More specifically, in the limit of $k\rightarrow0$, metric theories of gravity with energy-momentum conservation satisfy the adiabatic fluctuations of a flat universe with conservation of the comoving curvature $\zeta'=0$~\cite{weinberg:03,bertschinger:06}.
This has been shown to apply explicitly to the unified dark energy description used here in Ref.~\cite{gleyzes:14b}.
As a consequence, from momentum conservation it follows that
\bq
 \zeta''- \frac{H''}{H'}\zeta' = \Phi'' - \Psi' - \frac{H''}{H'}\Phi' - \left(\frac{H'}{H}-\frac{H''}{H'}\right)\Psi \rightarrow0. \label{eq:shevol}
\eq
Hence, in this limit, we can set $\fzeta=0$ and determine the evolution of the metric potentials from Eq.~(\ref{eq:shevol}) given a relation between $\Phi$ and $\Psi$ from the Einstein equations.
In $\Lambda$CDM and dark energy models, where $\gamma=1$, the evolution of the Newtonian potential obtained from Eq.~(\ref{eq:shevol}) applies on all scales and for the concordance case with $\mu=1$ also determines the matter density fluctuations and velocities.
Hence, for small deviations from $\Lambda$CDM, we expect that a pivot scale of $k_*=0$ provides a good approximation for $\fPhi$ and $\fzeta$ at $k>0$.

We can also directly derive an evolution equation for $\fPhi$ at $k\rightarrow0$, adopting Eq.~(\ref{eq:scaling}) to set $\fzeta\rightarrow0$ and $\Deltam\rightarrow0$ in the modified Einstein equations, Eqs.~(\ref{eq:semidyneinst1}) and (\ref{eq:semidyneinst2}), which is equivalent to Eq.~(\ref{eq:shevol}) supplemented with a modified Einstein equation.
This yields
\bqa
 \fPhi' + \fPhi^2 +  \left\{ \frac{H'}{H} + \frac{\bontwt}{\bont} + \frac{\alpha \bonsieit (\bont + \gnit) \frac{\rhom}{H^2 M^2}}{\bont \left[2 \alpha \bont \frac{H'}{H} - (\alpha \aH \eBfo + \aK \gni) \frac{\rhom}{H^2 M^2}\right]} \right\} \fPhi & & \nonumber\\
 + \frac{\bonfot}{\bont}  + \frac{\alpha (1 + \aT) \bonsieit \gon \frac{\rhom}{H^2 M^2}}{\bont \left[2 \alpha \bont \frac{H'}{H} - (\alpha \aH \eBfo + \aK \gni) \frac{\rhom}{H^2 M^2}\right]}
 & = & 0. \label{eq:fPhiSH}
\eqa
The unified dark energy modifications $\mu$ and $\gamma$ are then determined by Eq.~(\ref{eq:fPhiSH}), $\fzeta=0$, and Eqs.~(\ref{eq:mu}) and (\ref{eq:gamma}).
Note, however, since this implies $\fzeta=0$ on all scales, for models where $\fzeta$ contributes significantly at $k>0$, as we will encounter in Sec.~\ref{sec:examples}, setting the pivot scale at $k_*=0$ may not accurately recover the unified dark energy fluctuations.


\subsection{Choice of pivot scale} \label{eq:pivotchoice}

In the following, we discuss a few natural choices for the pivot scale $k_*$, for which the modified perturbed Einstein equations, Eqs.~(\ref{eq:einstein1}) and (\ref{eq:einstein2}), and energy-momentum conservation equations, Eqs.~(\ref{eq:momcon}) and (\ref{eq:encon}), can then be solved to obtain $\fPhi$ and $\fzeta$.
This determines the next-to-leading-order terms for the small-scale modifications introduced in Horndeski gravity and describes the leading-order modifications in beyond-Horndeski theories.
Depending on application, one may decide for different choices of pivot scale or use multiple pivot scales to improve the accuracy over a wide range of scales.
\begin{itemize}
 \item {\bf Super-Hubble scales:} As discussed in Sec.~\ref{sec:superH}, the comoving curvature is conserved at $k_*=0$.
Hence, in this limit we can adopt $\fzeta=0$ to simplify the modified Einstein equations and describe the evolution of $\fPhi$ through Eq.~(\ref{eq:fPhiSH}).
This approximation can, however, become inaccurate for models where $\fzeta$ contributes significantly at $k>0$.
 \item {\bf Hubble scale:} The Hubble scale describes the visible horizon and hence for observational purposes it is sufficient to restrict approximations to $k>aH$ and determine $\fPhi$ and $\fzeta$ from solving the modified Einstein and energy-momentum conservation equations at $k_*=H_0$.
 \item {\bf Sub-Hubble scales:} Given that close to the Hubble scale observations are limited by cosmic variance, one may also choose to solve the field equations at a sub-Hubble scale, e.g., $k_*=20H_0$, to improve the accuracy at the observationally more interesting scales, where measurements are more significant.
  For models where the sound speed of dark energy is significantly smaller than the speed of light, the contributions of velocity and time derivatives of the metric potentials at scales below the visible horizon can become important~\cite{sawicki:15}.
  In this case, a natural choice for the pivot scale would be the sound horizon $k_*=H_0/\cs$.
 \item {\bf Small scales:} Finally, the quasistatic approximation with $\fPhi=\fzeta=0$, applicable in Horndeski gravity in the limit of $k\rightarrow\infty$, allows one to determine the effective modifications from $\Lambda$CDM analytically from Eqs.~(\ref{eq:muinfH}) and (\ref{eq:gammainfH}) without the necessity of solving the field equations at a pivot scale, but it can break down near the Hubble scale. 
  Moreover, as pointed out in Sec.~\ref{sec:mod}, one should be cautious when applying this approximation when $\aH\neq0$, in which case $\fPhi$ and $\fzeta$ may give rise to leading-order contributions in $k$ to $\mu$ and $\gamma$.
  Hence, a sub-Hubble pivot scale may be of interest in particular for beyond-Horndeski models, with evaluation of $\fPhi$ and $\fzeta$ at $k_*\rightarrow\infty$.
\end{itemize}


\section{Examples in dark energy and modified gravity} \label{sec:examples}

We test the semi-dynamical approximation introduced in Sec.~\ref{sec:semidynamics} against numerical solutions of the exact modified linearly perturbed Einstein and energy-momentum conservation equations for the different free functions $\alpha_i(t)$ introduced in the unified dark energy formalism described in Sec.~\ref{sec:ude}.
For illustrative purpose, we set cosmological parameters to Planck 2015 values~\cite{planck13:15}.
In specific, we use the matter density parameter $\Om=0.308$ and dimensionless Hubble constant $h=0.678$, corresponding to $H_0\simeq2.26\times10^{-4}~h/{\rm Mpc}$.

Before studying effective modifications from concordance cosmology in the semi-dynamical approach, we quickly check that it consistently reproduces $\Lambda$CDM phenomenology in the corresponding limit.
In $\Lambda$CDM we have $\alpha_i=0$ $\forall i$ with $\kappa^2M^2=1$ and $H^2=H^2_{\Lambda{\rm CDM}}=H_0^2[\Om a^{-3}+(1-\Om)]$.
Using these relations in Eqs.~(\ref{eq:mu}) and (\ref{eq:gamma}) for $\mu$ and $\gamma$ with the coefficients defined in Appendix~\ref{sec:coefficients}, we obtain $\mu=\gamma=1$, irrespectively of the values for $\fPhi$ and $\fzeta$, as expected.
Hence, the metric potentials match, $\Phi=-\Psi$, and the standard Poisson equation relates the Newtonian potential to matter density fluctuations.
We can find the evolution of the fluctuations from solving
\bq
 \fPhi'+\fPhi^2+ \left(1 - \frac{H''}{H'}\right)\fPhi + \left(\frac{H'}{H} - \frac{H''}{H'}\right) = 0, \label{eq:LCDMevol}
\eq
which follows from Eqs.~(\ref{eq:shevol}) or (\ref{eq:fPhiSH}).

Next, we allow the contribution of a kinetic term and potential of the scalar field to the Einstein-Hilbert action in Sec.~\ref{sec:kinetic}, introducing $\aK\neq0$ and $H\neq H_{\Lambda{\rm CDM}}$, which represents the linear perturbation theory of quintessence models.
In Sec.~\ref{sec:mass}, we study metric $f(R)$ gravity as a representative example for models with a running Planck mass ($\aM\neq0$) and braiding ($\aB\neq0$).
Finally, in Sec.~\ref{sec:bh}, we alternatively allow a deviation of the speed of gravitational waves from the speed of light ($\aT\neq0$) and the introduction of beyond-Horndeski terms ($\aH\neq0$) to the action.
In all scenarios, we assume general relativistic initial conditions in the matter-dominated regime.


\subsection{Kinetic contribution and scalar field potential} \label{sec:kinetic}

We first study models with contribution of a scalar field kinetic energy ($\aK\neq0$) and potential, which allows for a free expansion history $H$.
This scenario is realised in quintessence models, which are represented by $\aK=3[1-\Om(a)][1+w(a)]$ with $w$ describing the equation of state of the dark energy component and $\aM=\aB=\aT=0$.
Thereby, Eqs.~(\ref{eq:mu}) and (\ref{eq:gamma}) become
\bq
 \mu = \frac{k_H^2}{-\left(\frac{2H'}{H} + \frac{\rhom}{H^2 M^2}\right)^{-2}\left[\left(\frac{2H'}{H} + \frac{\rhom}{H^2 M^2}\right) \left(\frac{\fzeta'}{\fzeta} + \fPhi\right) + \left(3 + \frac{H'}{H}\right) \frac{\rhom}{H^2 M^2} - \frac{2H''}{H}\right] \fzeta \aK + k_H^2}
\eq
and $\gamma=1$, respectively, where we have replaced $\fPhi'$ using $\zeta''=(\fzeta'+\fzeta\fPhi)\Phi$.
Hence, while there is no gravitational slip in quintessence models, we observe a clustering of the dark energy on large scales when $\fzeta\neq0$.
In the limit $k\rightarrow\infty$, we recover the quasistatic results $\mu=\gamma=1$, which thus do not account for the clustering effect at $k_H\sim1$.
Note that as we have $\fzeta=0$ in the limit of $k\rightarrow0$, setting the pivot scale at $k_*=0$ to extrapolate to sub-Hubble scales cannot account for the clustering either, hence, suggesting the choice of a sub-Hubble $k_*$.
The evolution of $\fPhi$ at $k\rightarrow0$, following from Eq.~(\ref{eq:shevol}) or (\ref{eq:fPhiSH}), is determined by Eq.~(\ref{eq:LCDMevol}) as in the concordance model.

\begin{figure}
 \centering
 \resizebox{\hsize}{!}{
  \resizebox{\hsize}{!}{\includegraphics{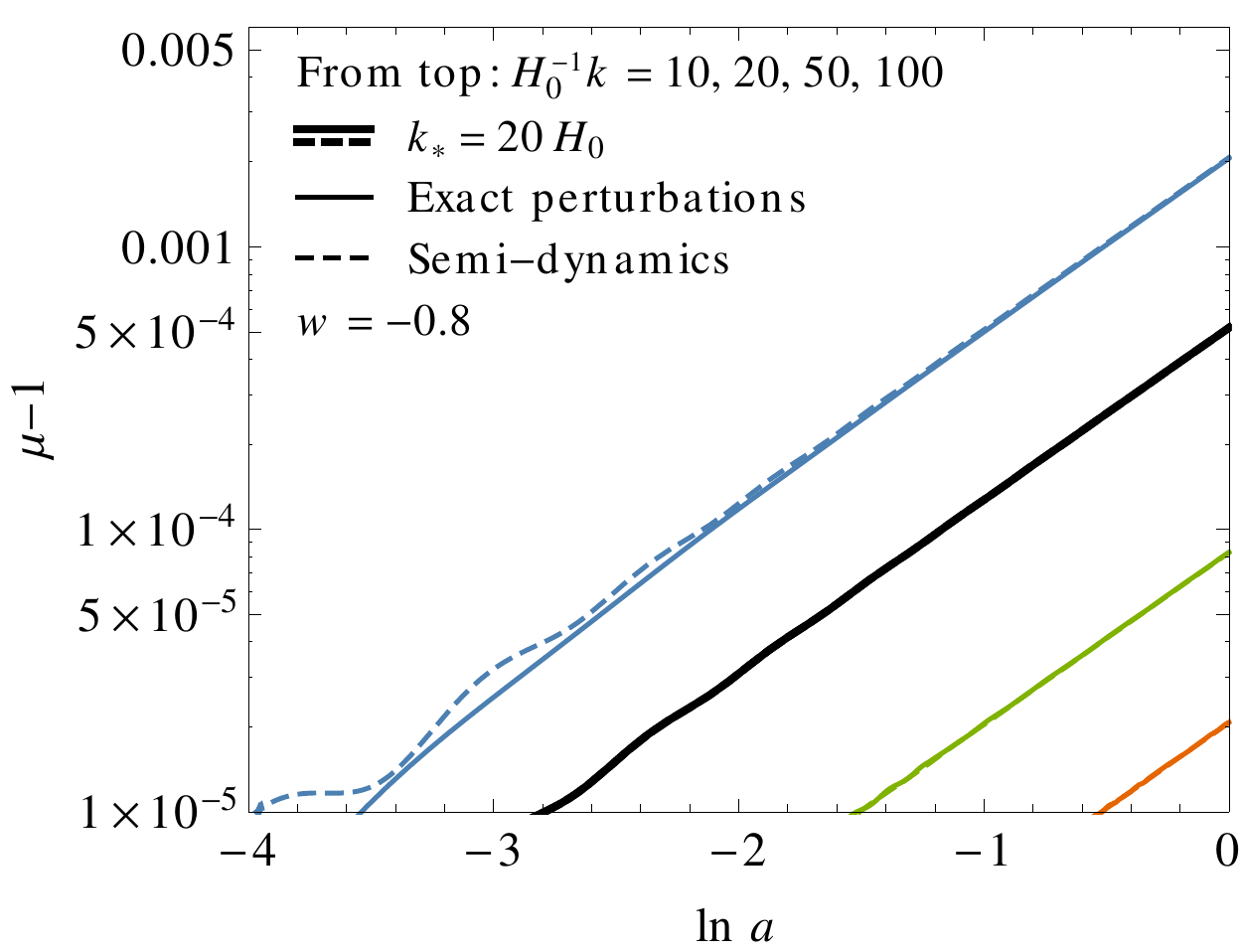}}
  \resizebox{0.963\hsize}{!}{\includegraphics{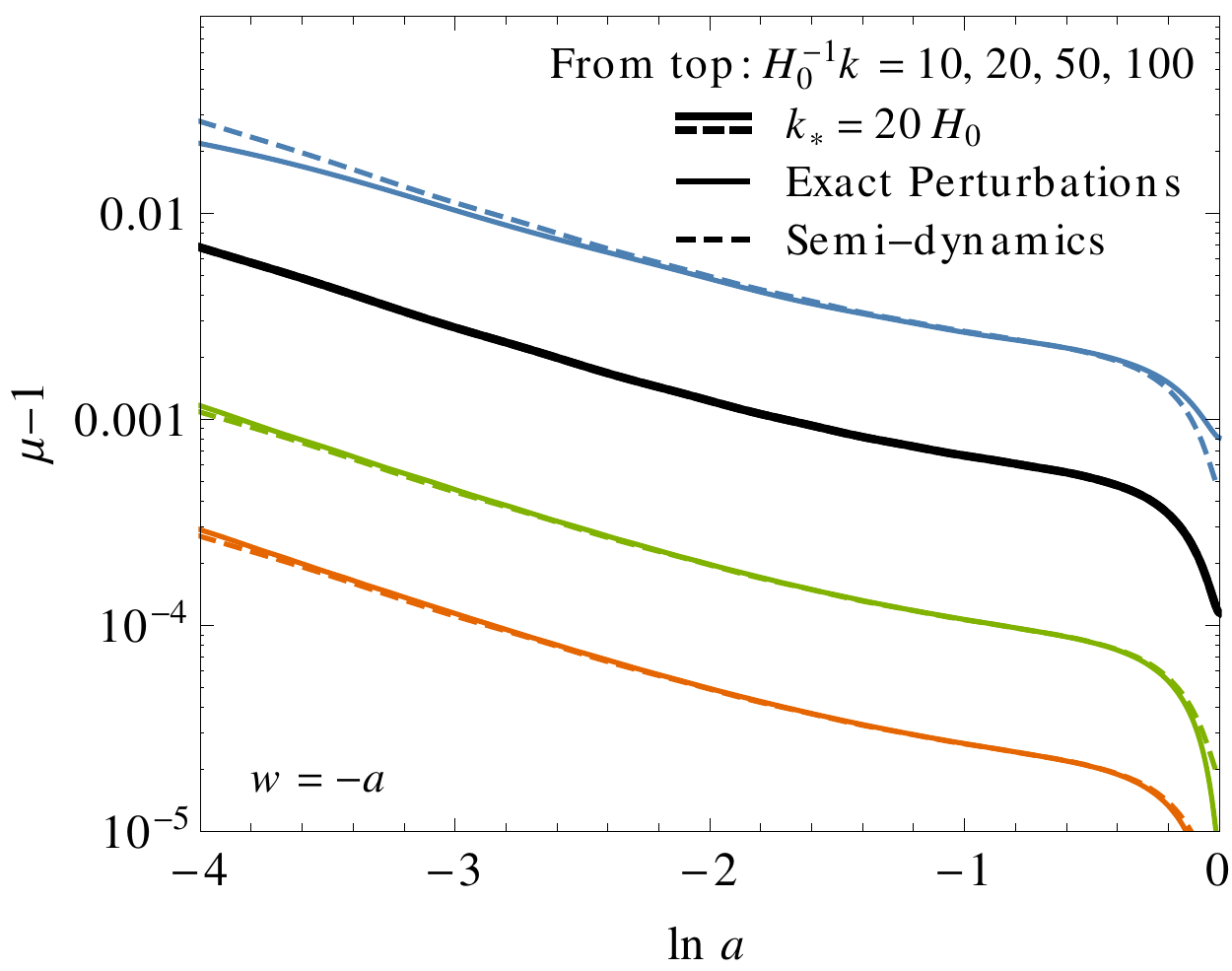}}
 }
\caption{Effective modification of the Poisson equation $\mu(a,k)$ determined from exact linear perturbation theory (solid lines) and the semi-dynamical approach (dashed lines) for a quintessence model with constant and time-dependent dark energy equation of state, $w=-0.8$ (left panel) and $w=-a$ (right panel), respectively.
A quasistatic approximation of the modified Einstein equations yields $\mu=1$ on all scales as in the concordance model.
The semi-dynamical approach introduces corrections to this result that match the exact perturbations of the dark energy models and describe the clustering of the quintessence field at very large scales.
We chose a pivot scale of $k_*=20H_0\simeq0.005~\hMpc$ to cover the observationally interesting scales of $k\gtrsim10H_0$.
\label{fig:quintessence}}
\end{figure}

In Fig.~\ref{fig:quintessence}, we show numerical results for the semi-dynamical approximation of the modification of the Poisson equation $\mu(a,k)$ for quintessence models with dark energy equation of state $w=-0.8$ and $w=-a$.
Thereby, we chose a pivot scale of $k_*=20H_0$, which accurately recovers the modifications at the observationally interesting scales of $k\gtrsim10H_0$ that are determined from solving the exact perturbed modified Einstein and energy-momentum conservation equations~\cite{lombriser:13a}.
Finally, note that for the models studied here, due to the weak scale dependence in $\fPhi$ and $\fzeta$ the logarithmic growth rate $\fDelta$ can directly be computed from Eq.~(\ref{eq:fDelta}) to good accuracy, as we have checked numerically.

\subsection{Evolving Planck mass and braiding} \label{sec:mass}

A running of the Planck mass is described by $\aM\neq0$ whereas braiding or mixing of the metric and scalar kinetic terms is described by $\aB\neq0$.
A scenario in which both parameters are nonzero is, for instance, realised in $f(R)$ models, where the free nonlinear function $f(R)$ is added to the Einstein-Hilbert action~\cite{buchdahl:70}, and represented by $\aM=2\aB=f_R'/(1+f_R)$ and $\aK=\aT=0$ with subscripts of $R$ denoting derivatives with respect to the Ricci scalar.
In $f(R)$ gravity $\cs^2=1$, which can be determined from Eq.~(\ref{eq:soundspeed}) using the Friedmann equations that can be found in Refs.~\cite{gubitosi:12,bloomfield:12,bellini:14,lombriser:14b,gleyzes:14b}.
This simplifies expressions for $\mu$ and $\gamma$ to
\bq
 \mu(a,k) = \frac{1}{\kappa^2M^2}\frac{\mu_{+2} k_H^2 + \frac{2}{3}\aM k_H^4}{\mu_{-0} + \mu_{-2}k_H^2 + \frac{1}{2}\aM k_H^4}, \ \ \ \ \ \ \gamma(a,k) = \frac{\gamma_{+0} + \frac{1}{3}\aM k_H^2}{\mu_{+2} + \frac{2}{3}\aM k_H^2}, \label{eq:modfR}
\eq
where $\kappa^2M^2=1+f_R$ and the coefficients of Eqs.~(\ref{eq:modfR}) are given by 
\bqa
 \mu_{+2} & = & \aM \left[\aM + \fPhi' + \fPhi^2 + \left(3 +  \frac{H'}{H}\right) \fPhi - 1 + (\fzeta - \fPhi) \frac{H}{H'} \frac{\rhom}{H^2 M^2}\right] \nonumber\\
  & & + \frac{H'}{H} \left(4 + \frac{H'}{H} + \frac{H''}{H'}\right), \\
 \mu_{-0} & = & \frac{3}{2} \aM \left\{\aM \left[\aM (2 \fPhi + 1) + \fPhi' + \fPhi^2 + (1 - \fPhi) \frac{H'}{H} - 1 + \frac{H}{H'} \fzeta \frac{\rhom}{H^2 M^2}\right] \right. \nonumber\\
  & & \left. - \frac{H'}{H} \left[\fPhi' + \fPhi^2 - \left(1 + \frac{H''}{H'}\right) \fPhi + \frac{H'}{H} - \frac{H''}{H'} - 2 + \left(\fzeta - \fPhi - 1\right) \frac{H}{H'} \frac{\rhom}{H^2 M^2}\right]\right\}, \\
 \mu_{-2} & = & \frac{1}{2} \left\{\aM \left[\aM (\fPhi + 5) + \fPhi' + \fPhi^2 + \left(3+\frac{H'}{H}\right)\fPhi - 3\frac{H'}{H} - 4 \right. \right. \nonumber\\
  & & \left. \left. + \frac{H}{H'}(\fzeta - \fPhi + 1) \frac{\rhom}{H^2 M^2}\right] + 2\frac{H'}{H} \left(4 + \frac{H'}{H} + \frac{H''}{H'}\right)\right\}, \\
 \gamma_{+0} & = & \aM \left(2\aM - 3 - \frac{2 H'}{H} + \frac{H}{H'} \frac{\rhom}{H^2 M^2}\right) + \frac{H'}{H} \left(4 + \frac{H'}{H} + \frac{H''}{H'}\right).
\eqa
In the limit of $k\rightarrow\infty$, we recover $\mu_{\infty}=(1+f_R)^{-1}4/3$ and $\gamma_{\infty}=1/2$.

\begin{figure}
 \centering
 \resizebox{\hsize}{!}{
  \resizebox{\hsize}{!}{\includegraphics{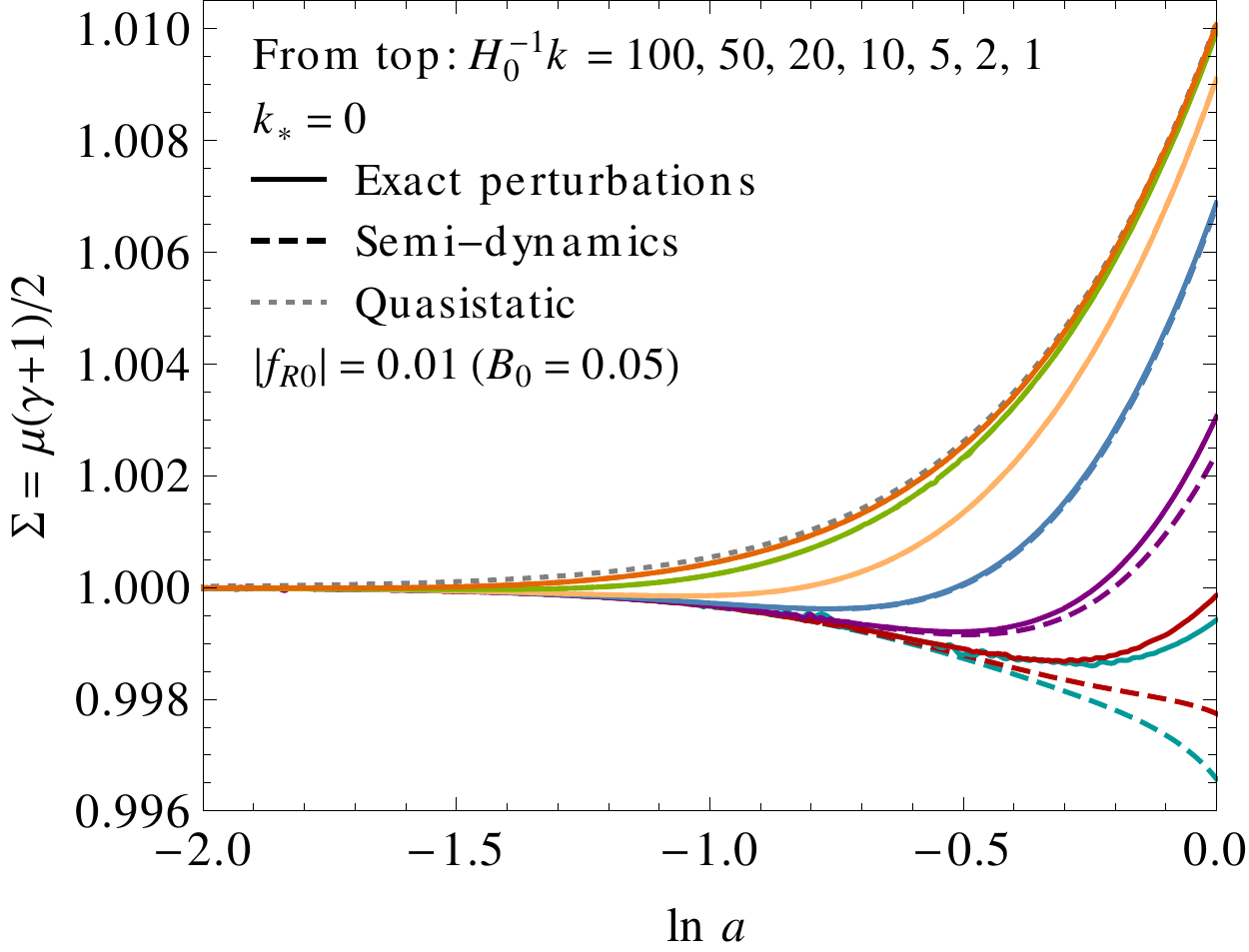}}
  \resizebox{0.945\hsize}{!}{\includegraphics{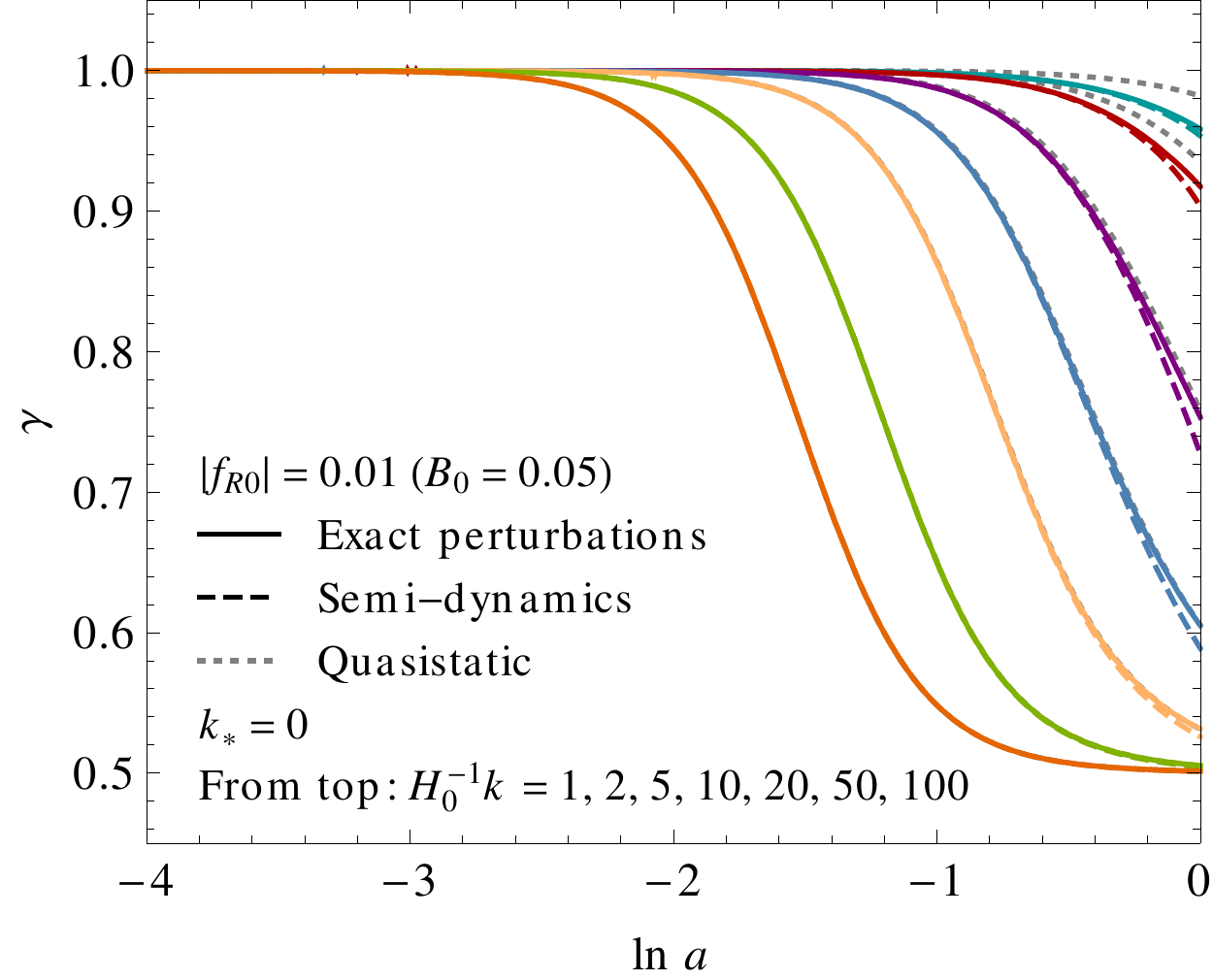}}
 }
\caption{Effective modification of the relation of the lensing potential $(\Phi-\Psi)/2$ to matter density fluctuations $\Sigma(a,k)=\mu(\gamma+1)/2$ (left panel) and the gravitational slip $\gamma(a,k)$ (right panel) determined from exact linear perturbation theory (solid lines), the semi-dynamical approach (dashed lines), and the quasistatic approximation (dotted lines) for $f(R)$ gravity with $|f_{R0}|=0.01$ ($B_0\simeq0.05$).
The pivot scale is set at $k_*=0$.
Note that a conventional quasistatic approximation yields a scale-independent $\Sigma=(1+f_R)^{-1}$, corresponding to the $k\rightarrow\infty$ limit ($k\approx100H_0$) in the semi-dynamical approach, and hence does not account for the scale dependence in $\Sigma$ of the exact solution.
\label{fig:fR}}
\end{figure}

To test the performance of the semi-dynamical approximation against the exact perturbations of $f(R)$ gravity~\cite{lombriser:10,song:06}, we set the pivot scale at $k_*=0$, where $\fzeta=0$ and solve
\bq
 \fPhi'+\fPhi^2 + \left(1+\frac{\aM^2 - \aM' - 2 \aM \frac{H'}{H} + \frac{H''}{H}}{\aM - \frac{H'}{H}}\right)\fPhi - \frac{\aM' + \left(\frac{H'}{H}\right)^2 - \frac{H''}{H}}{\aM - \frac{H'}{H}} = 0, \label{eq:fRSH}
\eq
which follows from Eq.~(\ref{eq:fPhiSH}).
Note that Eq.~(\ref{eq:fRSH}) is equivalent to the super-Hubble evolution found in Eq.~(43) of Ref.~\cite{song:06} with $\aM=H'B/H$, where the Compton wavelength parameter is defined as $B\equiv f_{RR}(1+f_R)^{-1}R'H/H'$.
The resulting effective modifications are shown in Fig.~\ref{fig:fR} for a designer $f(R)$ gravity model~\cite{song:06} with $|f_{R0}|=0.01$ ($B_0\simeq0.05$) and an equation of state of the dark energy component of $w=-1$, where subscripts of zero denote evaluation today.
These parameter values are cosmologically not viable~\cite{lombriser:14a} and serve here only for illustration.
To better illustrate deviations from the quasistatic approximation, we show $\Sigma\equiv\mu(\gamma+1)/2$ rather than $\mu$ in Fig.~\ref{fig:fR}, where $\Sigma$ relates the lensing potential $(\Phi-\Psi)/2$ to the matter density fluctuations using Eqs.~(\ref{eq:mu}) and (\ref{eq:gamma}).
In the conventional quasistatic limit, neglecting velocities and time derivatives of $\Phi$ and $\Psi$ in the field equations in Appendix~\ref{sec:einsteineqs}, one obtains a scale-independent $\Sigma=(1+f_R)^{-1}$ whereas the semi-dynamical approach introduces a scale dependency in $\Sigma$, however, with the effects being too small to be observed with lensing~\cite{lombriser:11b}.
Finally, note that for $f(R)$ models, due to the strong scale dependence in $\fPhi$ and $\fzeta$, determining the logarithmic growth rate $\fDelta$ directly from Eq.~(\ref{eq:fDelta}) does not yield a good approximation and one should integrate Eq.~(\ref{eq:Fcal}) to obtain more accurate results.


\subsection{Varying tensor speed and beyond-Horndeski terms} \label{sec:bh}

\begin{figure}
 \centering
 \resizebox{\hsize}{!}{
  \resizebox{\hsize}{!}{\includegraphics{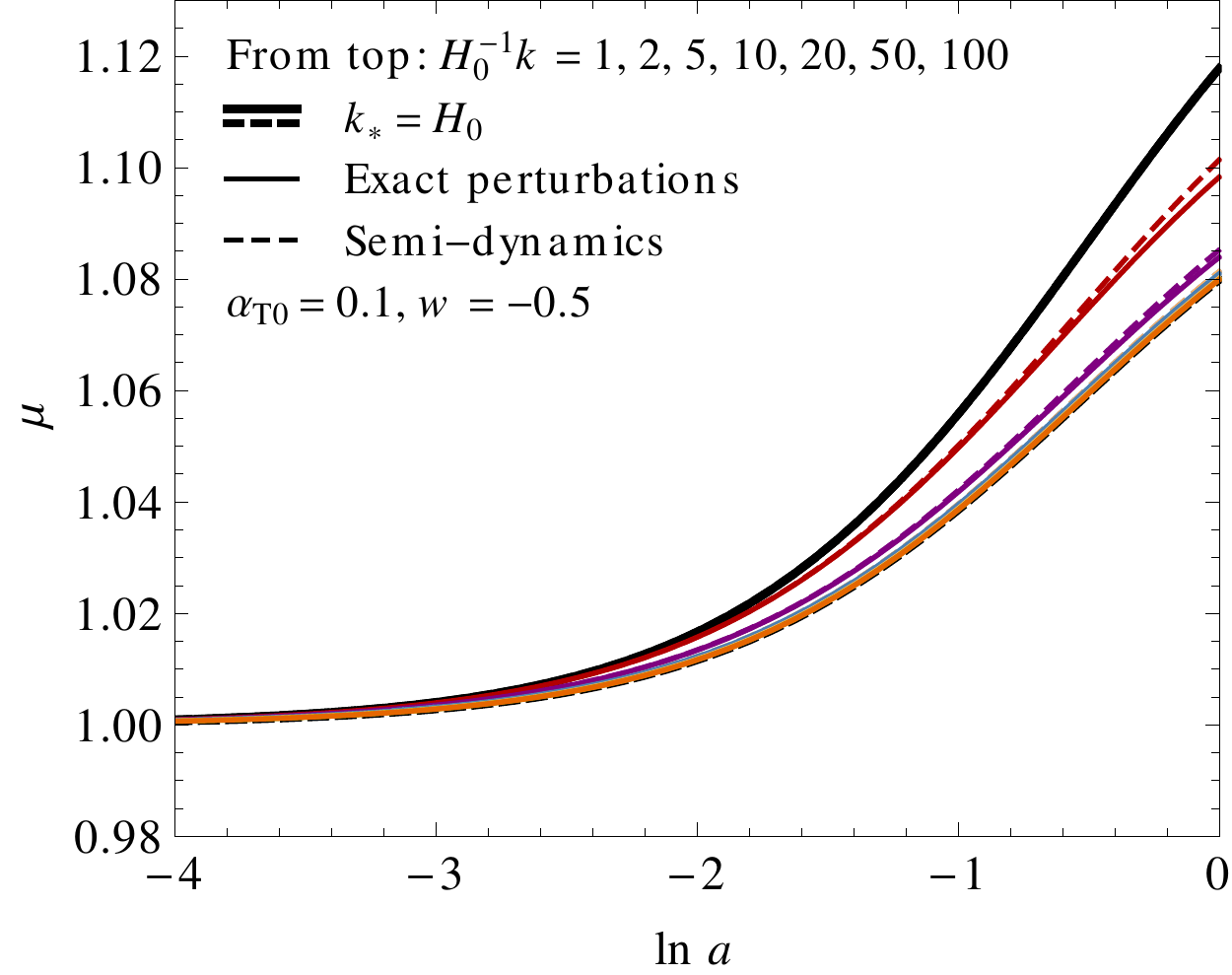}}
  \resizebox{\hsize}{!}{\includegraphics{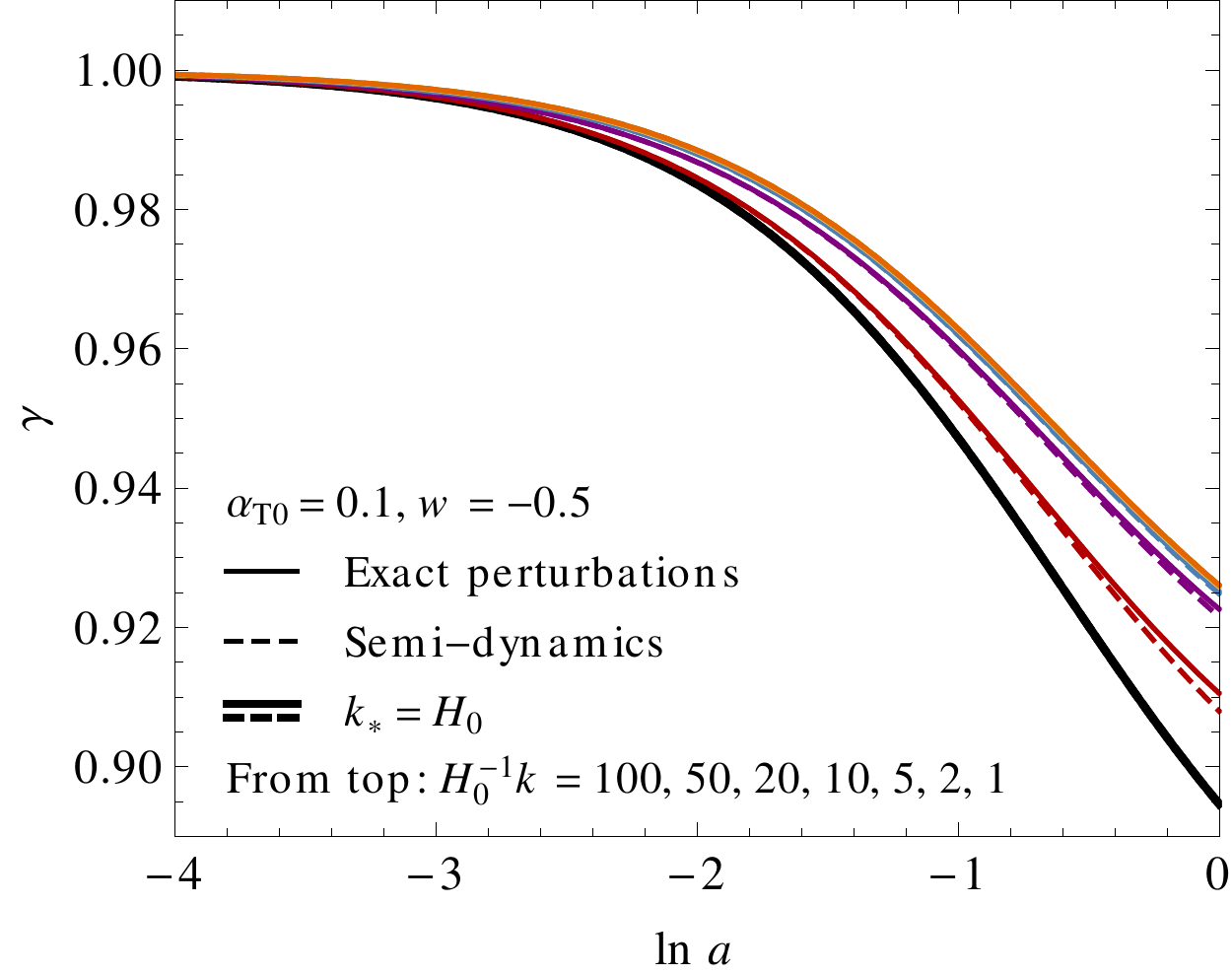}}
 }
\caption{Effective modification of the Poisson equation $\mu(a,k)$ (left panel) and gravitational slip $\gamma(a,k)$ (right panel) determined from exact linear perturbation theory (solid lines) and the semi-dynamical approach (dashed lines) for a modification of the tensor speed with $\aT\neq0$.
The pivot scale is set to $k_*=H_0$.
\label{fig:tensorspeed}}
\end{figure}

Finally, we study scenarios in which alternatively $\aT\neq0$ or $\aH\neq0$.
We start with the tensor speed modification and set $\aT=\alpha_{{\rm T}0}[1-\Om(a)]$.
With a concordance model expansion history, we obtain $\gamma=k_H^2/(3\aT\fzeta+k_H^2)$ and $\mu=1+3\aT\fzeta/k_H^2$.
Hence, for $\fzeta\ll1$ deviations from the concordance model structure remain small at sub-Hubble scales.
However, the expressions for $\gamma$ and $\mu$ become more complicated when $H$ deviates from its $\Lambda$CDM evolution.
For illustrative purpose, we set $w=-1/2$ and $\alpha_{{\rm T}0}=0.1$ and solve the modified Einstein equations in Appendix~\ref{sec:einsteineqs} for $\kappa^2M^2=1$ and $\alpha_i=0$ ($i\neq{\rm T}$).
The super-Hubble evolution follows from Eq.~(\ref{eq:fPhiSH}), which here becomes
\bq
 \fPhi' + \fPhi^2 + \left(1 - \frac{\aT' + \frac{H'}{H}\aT + \frac{H''}{H}}{\aT + \frac{H'}{H}}\right)\fPhi + \frac{\aT' \left(\frac{H'}{H}-1\right) + (1 + \aT) \left[\left(\frac{H'}{H}\right)^2 - \frac{H''}{H}\right]}{\aT + \frac{H'}{H}} = 0.
\eq
However, since $\fzeta=0$ at $k\rightarrow0$, we set $k_*$ at the Hubble scale.
The corresponding results are shown in Fig.~\ref{fig:tensorspeed}, where we find good agreement between the semi-dynamical approximation and the exact modifications.
We also find that if choosing a pivot scale at sub-Hubble scales for the model adopted here, Eq.~(\ref{eq:fDelta}) provides a good approximation to the logarithmic growth rate $\fDelta$ since $\fPhi$ and $\fzeta$ are only weakly dependent on scale within the visible horizon.

\begin{figure}
 \centering
 \resizebox{\hsize}{!}{
  \resizebox{\hsize}{!}{\includegraphics{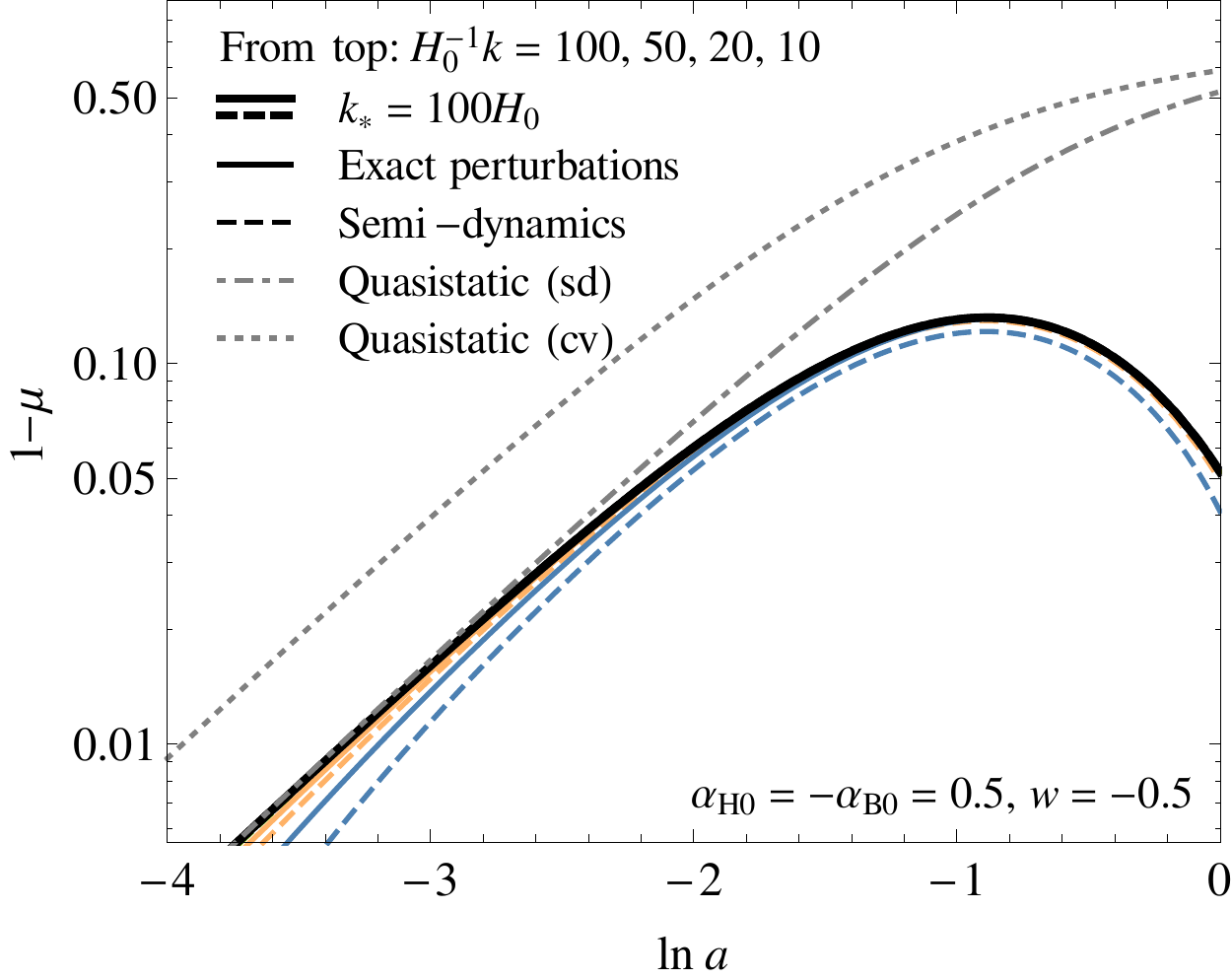}}
  \resizebox{\hsize}{!}{\includegraphics{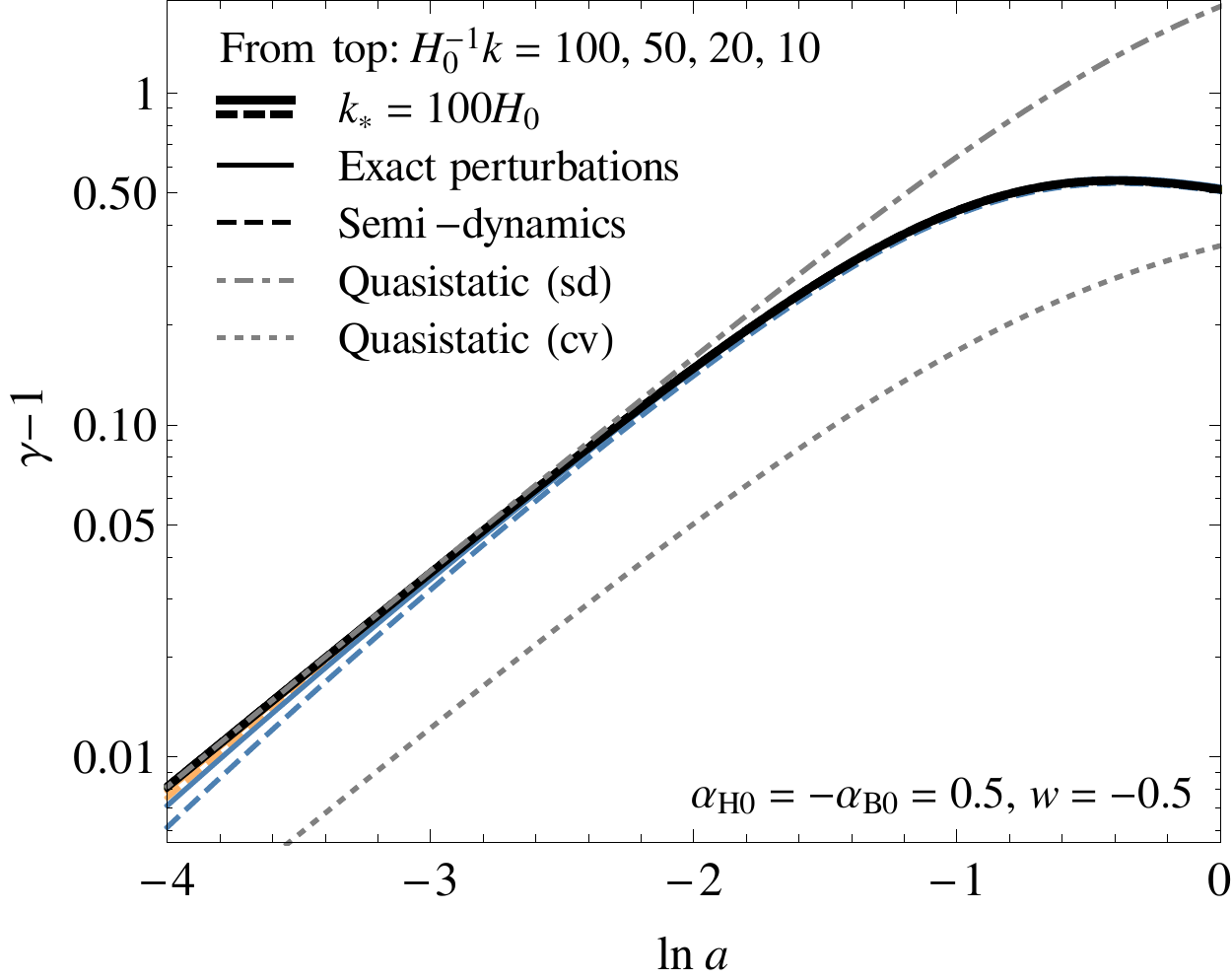}}
 }
\caption{Effective modification of the Poisson equation $\mu(a,k)$ (left panel) and the gravitational slip $\gamma(a,k)$ (right panel) determined from exact linear perturbation theory (solid lines), the semi-dynamical approach (dashed lines), and quasistatic approximations for a modification with beyond-Horndeski term $\aH\neq0$.
The pivot scale is set to $k_*=100H_0$ ($k\rightarrow\infty$).
The two quasistatic approximations are obtained by setting $\fPhi=\fzeta=0$ in the semi-dynamical (sd; dot-dashed curve) equations with $k=100H_0$ and by a conventional approach (cv; dotted curves) of neglecting $\Vm/k_H$ with respect to $\Deltam$ and time derivatives of the fields in the modified Einstein equations in Appendix~\ref{sec:einsteineqs} in the limit of $k\rightarrow\infty$.
The velocity field and the time derivative of the metric potential $\Phi$ contribute to $\mu$ and $\gamma$ at leading order in $k$, which prevents the quasistatic approximations to reproduce the exact perturbation results.
\label{fig:beyondH}}
\end{figure}

Finally, we analyse the new phenomenological aspects that beyond-Horndeski theories~\cite{gleyzes:14a} introduce with $\aH\neq0$.
Thereby we adopt $\aH=\alpha_{{\rm H}0}[1-\Om(a)]$ and require $\kappa^2M^2=1$ and $\aM=\aK=\aT=0$.
In order to have a well-defined general relativistic limit $\mu=\gamma=1$ when $\aH\rightarrow0$, we set $\aB=-\aH$.
For illustrative purpose, we set $\alpha_{{\rm H}0}=1/2$ and also adopt $w=-1/2$ as in the case of $\aT\neq0$, and solve the modified Einstein equations in Appendix~\ref{sec:einsteineqs} in the corresponding limit.
The super-Hubble evolution of $\fPhi$ reduces to Eq.~(\ref{eq:LCDMevol}).
As discussed in Sec.~\ref{sec:semidynamics}, in case of nonzero $\aH$, $\Phi'$ and $\Vm/k_H$ contribute to $\gamma$ and $\mu$ through $\fPhi$ and $\fzeta$ at leading order in $k$.
Hence, setting the pivot scale at $k_*=0$, where $\fzeta=0$, does not account for the full correction.
In Fig.~\ref{fig:beyondH}, we compare the modifications $\mu$ and $\gamma$ against the semi-dynamical approximation using the pivot scale $k_*=100H_0$ ($k\rightarrow\infty$) instead, finding excellent agreement between the two for scales of $k\gtrsim10H_0$.
To illustrate the importance of nonzero contributions of $\fPhi$ and $\fzeta$, we also show the resulting modifications at $k=100H_0$ derived in the quasistatic approximation of setting $\fPhi=\fzeta=0$ in Eqs.~(\ref{eq:mu}) and (\ref{eq:gamma}).
Moreover, we show results from a more conventional quasistatic approach of assuming domination of $\Deltam$ over $\Vm/k_H$ and neglecting time derivatives of the fields in the modified Einstein equations given in Appendix~\ref{sec:einsteineqs} in the limit of $k\rightarrow\infty$, which yields
\bq
 \mu^{-1}=(1+\aH)^2+\aH\frac{2\aH'+2(\aH-1)\left(\frac{H'}{H}+3\right)+\frac{\kappa^2\rhom}{H^2}}{\frac{H'}{H}\left(\frac{H'}{H}+\frac{H''}{H'}+3\right)}
\eq
and $\gamma=1+\aH$.
In both approximations, there are $\mathcal{O}(1)$ deviations from their counterparts computed from the exact perturbations, as expected (see Fig.~\ref{fig:beyondH}).
Note, however, that the quasistatic approximation within the semi-dynamical approach, setting $\fPhi=\fzeta=0$, performs well at high redshifts, where $|\aH|\ll1$, which is in agreement with a similar analysis performed in Refs.~\cite{defelice:15,tsujikawa:15}.
Finally, we find that due to the weak scale dependence in $\fPhi$ and $\fzeta$, Eq.~(\ref{eq:fDelta}) provides a good approximation to the logarithmic density growth rate $\fDelta$ in the scenario discussed here.


\section{Discussion} \label{sec:conclusion}

The development of a generalised description of the cosmological structure formed in modified gravity and dark energy models has been a very active field of research in recent years.
One such formalism resulting from this exploration is the effective field theory of cosmic acceleration or unified dark energy~\cite{gubitosi:12,bloomfield:12}, encompassing Horndeski scalar-tensor theories~\cite{horndeski:74} and beyond~\cite{gleyzes:14a}.
While the full system of differential equations in the EFT framework can be solved numerically, this can be computationally challenging as well as inefficient depending on application.
Moreover, the model-specific solutions may only offer little insight into the generic behaviour of the wide range of modifications possible.
Hence, a quasistatic approximation, neglecting time derivatives with respect to spatial derivatives, is often applied to simplify computations.
This allows one to extract information about the relations between the fluctuations without the need of integrating the field and conservation equations.
The approximation can, however, break down near the Hubble scale or the sound horizon for models with subluminal sound speed of the dark energy, affecting tests of gravity and dark energy employing observations at very large scales~\cite{lombriser:13a,duniya:13,sawicki:15}.
Furthermore, an analysis beyond the quasistatic regime becomes necessary to constrain the entire available EFT model space and break degeneracies with $\Lambda$CDM for particular subsets of this space~\cite{lombriser:14b}, hence, demanding a more general, yet efficient approach.

Here, we have developed a semi-dynamical treatment of the linear cosmological perturbations of unified dark energy that accounts for the time derivatives of the metric potentials and large-scale velocity fields. Our approximation extrapolates the evolution of fluctuations at a pivot scale to arbitrary scales and facilitates the analysis of modifications from the concordance model in its application to observational tests of gravity and dark energy.
We adopted the unified dark energy formalism of Ref.~\cite{bellini:14,gleyzes:14b} and derived the energy-momentum conservation equations and a set of combined modified Einstein equations, eliminating the scalar field fluctuations, for the fully available model space, including beyond-Horndeski models.
Evaluating the velocity field and rate of growth of the spatial metric potential at the pivot scale, we determined the time and scale dependence of the effective modifications from the concordance model in the form of a deviation in the Poisson equation and a gravitational slip and approximated the growth rate of matter density fluctuations.
At small scales, and for Horndeski theories, the resulting modifications recover a quasistatic approximation while accounting for corrections at the next-to-leading order in $k$, which become important near the Hubble scale.
For models beyond Horndeski gravity, we found that the velocity field and time derivative of the spatial metric potential can generally not be neglected and contribute at first order even in the small-scale limit.
Finally, we tested the semi-dynamical approximation against numerical solutions of the exact modified Einstein and energy-momentum conservation equations for a range of dark energy and modified gravity models, including kinetic contributions of the scalar field, running Planck mass, kinetic braiding or mixing, deviations between the speed of gravitational waves and the speed of light, and beyond-Horndeski terms.
We found good agreement between the approximations and the exact results.

Since we only require the integration of the field equations at a single scale, the semi-dynamical approximation can be used to significantly enhance the efficiency in numerically intensive observational tests of gravity and dark energy, or to perform analytical operations that help to gain more insights on the rich phenomenology of unified dark energy.


\section*{Acknowledgements}
The authors thank Pedro Ferreira and Kazuya Koyama for useful discussions and comments.
This work has been supported by the STFC Consolidated Grant for Astronomy and Astrophysics at the University of Edinburgh.
Please contact the authors for access to research materials.


\newpage
\appendix


\section{Modified perturbed Einstein equations}\label{sec:einsteineqs}

We present here the modified perturbed Einstein equations in the unified dark energy formalism of Refs.~\cite{bellini:14,gleyzes:14b} that we have used in Sec.~\ref{sec:semidynamics} to describe the effective modifications of the Poisson equation $\mu(a,k)$ and the gravitational slip $\gamma(a,k)$.
Metric perturbations are described in the Newtonian gauge with the FLRW line element given in Eq.~(\ref{eq:FLRW}).
However, contrary to Refs.~\cite{bellini:14,gleyzes:14b}, we adopt the total matter gauge for the matter fluctuations.

With these specifications, the time-time component of the modified perturbed Einstein equation, or Hamiltonian constraint, becomes
\bqa
 & & 6(\aB+1)\Phi' + 2(\aH+1)k_H^2\Phi - (12\aB-\aK+6)\Psi + (6\aB-\aK) H\pi' \nonumber\\
 & & - 2\left[(\aH-\aB)k_H^2 + 3(\aB+1)\frac{H'}{H} + \frac{3}{2}\frac{\rhom}{H^2 M^2}\right] H\pi = \frac{\rhom}{H^2 M^2}\left(\Delta-3\frac{\Vm}{k_H}\right). \label{eq:timetime}
\eqa
The time-space component, or momentum constraint, is
\bqa
\Phi' - (\aB+1)\Psi + \aB H\pi' - \left(\frac{H'}{H} + \frac{\rhom}{2 H^2 M^2}\right) H\pi & = & -\frac{\rhom}{2 H^2 M^2}\frac{\Vm}{k_H}. \label{eq:timespace}
\eqa
The traceless space-space component, or anisotropy constraint, is given by
\bqa
 (\aT+1)\Phi + (\aH+1)\Psi - \aH H\pi' + (\aM-\aT)H\pi & = & 0 \label{eq:tracelessspacespace}
\eqa
and the trace of the space-space component, or pressure equation, by
\bqa
 & & \Phi'' + \left(\aM+\frac{H'}{H}+3\right)\Phi'- (\aB+1)\Psi' - \left[\aB' + (\aB+1) \left(\aM+\frac{H'}{H}+3\right) - \frac{\rhom}{2 H^2 M^2}\right]\Psi \nonumber\\
 & & + \aB H\pi'' + \left[\aB' + \aB\left(\aM + \frac{2H'}{H}+3\right) - \frac{H'}{H} - \frac{\rhom}{2 H^2 M^2} \right]H\pi' \nonumber\\
 & & - \left[\left(\aM + \frac{H'}{H} + 3\right)\frac{H'}{H} + \frac{H''}{H}\right] H\pi = 0. \label{eq:tracespacespace}
\eqa
Finally, the scalar field equation becomes
\bqa
 & & \aK H\pi'' + \left\{\aK' + \aK\left[\aM+3\left(\frac{H'}{H}+1\right)\right]\right\} H\pi' \nonumber\\
 & & - 2 \left\{\left[\aB'-\aH'-\aM+\aT + (\aB-\aH)\left(\aM+\frac{H'}{H}+1\right)+\frac{H'}{H} + \frac{\rhom}{2 H^2 M^2}\right] k_H^2 \right. \nonumber\\
 & & \left. - 3 \frac{H'}{H}\left[\aB' + \aB \left(\aM + \frac{H''}{H'} + 2\frac{H'}{H} + 3\right) + \frac{H'}{H} + \frac{\rhom}{2 H^2 M^2}\right]\right\} H\pi - 6\aB\Phi'' \nonumber\\
 & & - 2\left\{ \aH k_H^2 + 3 \left[ \aB' + \aB \left(\aM + 2\frac{H'}{H} + 3\right) + \frac{H'}{H} + \frac{\rhom}{2 H^2 M^2}\right]\right\}\Phi' + (6\aB-\aK)\Psi' \nonumber\\
 & & + 2\left[- \aH' - \aM + \aT - \aH(\aM + 1)\right]k_H^2\Phi + 2\left[ (\aH - \aB)k_H^2 + \left(3\aB - \frac{\aK}{2}\right)(\aM + 3) \right. \nonumber\\
 & & \left. + \left(3\aB'-\frac{\aK'}{2}\right) + (9\aB-\aK+3)\frac{H'}{H} + \frac{3}{2}\frac{\rhom}{H^2 M^2}\right]\Psi = 0. \label{eq:scalarfieldeq}
\eqa


\newpage
\section{Coefficients}\label{sec:coefficients}

Finally, we provide a list of the coefficients appearing in the combined perturbed modified Einstein equations, Eqs.~(\ref{eq:einstein1}) and (\ref{eq:einstein2}), in Sec.~\ref{sec:ude} and contributing to the semi-dynamical approximation of $\mu(a,k)$ and $\gamma(a,k)$ in Eqs.~(\ref{eq:mu}) and (\ref{eq:gamma}), respectively, in Sec.~\ref{sec:mod}.
Note that the parameters $\beta_i$ and $\gamma_i$, which are independent of $\aH$ since defined for Horndeski models, are provided in Refs.~\cite{bellini:14,gleyzes:14b} and are not given here again.

\paragraph{Combined modified Einstein equations} \label{sec:combeinsteineqs}

The beyond-Horndeski terms ($\aH\neq0$) in the combined field equations and the modifications $\mu$ and $\gamma$ are defined as follows:
\bqa
 \en & = & -\frac{2}{\alpha} (\aB - \aH)^3, \\
 \eon & = & -3 \aB \frac{\rhom}{2 H^2 M^2}, \\
 \etw & = & (\beta_3 + \epsilon_{4a}) \eon, \\
 \ethr & = & -\frac{2}{\aB - \aH} \left( \frac{\aB'}{\aB} - \frac{\aH'}{\aH} + \frac{\rhom}{4 H^2 M^2} \right), \\
 \efo & = & \left[\beta_4 - (1 + \aT) \left(\beta_2 - \frac{\etw}{\eon}\right)\right] \eon, \\
 \efi & = & (\aB - \aH - 1) \left[\left(\frac{H'}{H}\right)^2 + \frac{H''}{H}\right] + \frac{\aH + 1}{4 \alpha} (12 \aB - \aK - 6 \aH) \left(\frac{\rhom}{H^2M^2}\right)^2 \nonumber\\
  & & + \frac{1}{2} \left[ \frac{\epsilon_{5a}}{\alpha} - \aB (\aM - \beta_3 + 3) - 2 (\aM - \aT - \aH') - \aH (\aM + 1) + \frac{\aH'}{\aH} \right] \frac{\rhom}{H^2M^2} \nonumber\\
  & & + \left\{ \epsilon_{5b} + (\aH + 1) \left[ \frac{1}{2} + \frac{1}{\alpha} \left(6\aB - \aK - 3 \aH \{\aB + 1\} \right) \right] \frac{\rhom}{H^2M^2} \right\} \frac{H'}{H} + \epsilon_{5c}, \\
 \esi & = & \beta_1 \left\{ \ese - \frac{2}{\alpha} \left[(\aB - \aH) \ethr + \beta_2 - \beta_3\right]\right\} - 2 \left\{(1 + \aB) \left[\left(2 - \frac{\aH}{\aB}\right) \frac{H'}{H} - \aM + \aT\right] \right. \nonumber\\
  & & \left. - \aH \aB \left(\frac{\aB'}{\aB^2} - \frac{\aH'}{\aH^2}\right) + \frac{2 \aB - \aH}{\aB} \frac{\rhom}{2 H^2 M^2}\right\} \frac{\eon}{\alpha}, \\
 \ese & = & \frac{2 \aB}{\alpha} \left[2 \left(\frac{\aB'}{\aB} - \frac{\aH'}{\aH}\right) - \epsilon_{4a} + \frac{\aH}{\aB} \frac{\rhom}{2 H^2 M2} \right], \\
 \eonsiei & = & 3 \left[\frac{2 \aB' \aK - \aB \aK'}{\alpha} - \aB \left(\epsilon_{4a} - \aM + \frac{H'}{H} + 2\right)\right] \left(\aM - \aT - \frac{H'}{H}\right) \nonumber\\
  & & - 3\aB \left(\aM^2 - \aM \aT + \aM' - \aT' - \aT \frac{H'}{H} - \frac{H''}{H}\right), \\
 \eseni & = & \frac{6 \aB}{\alpha} \left[2 \left(\frac{\aB'}{\aB} - \frac{\aH'}{\aH}\right) - \epsilon_{4a}\right] - \frac{1}{\aB - \aH} \left[2 \left(\frac{\aB'}{\aB} - \frac{\aH'}{\aH}\right) + \aM - \aT - \frac{H'}{H}\right], \\
 \eBon & = & (1 + \aH) \left( \frac{H'}{H} + \frac{\rhom}{2 H^2 M^2}\right) - \aB  (1 + \aM), \\
 \eBthr & = & \frac{1}{2} \left( \frac{H'}{H} + \frac{\rhom}{2 H^2 M^2}\right), \\
 \eBfo & = & -3 \aB \frac{H'}{H},
\eqa
\bqa
 \epsilon_{4a} & = & \frac{\aB'}{\aB} - \frac{\aH'}{\aH} + 1 + \aM - \frac{H'}{H} - \frac{1}{\aB} \left(\frac{H'}{H} + \frac{\rhom}{2 H^2 M^2}\right), \\
 \epsilon_{5a} & = & 6 \aB \aH \left[ \left(\aM - \frac{\aB'}{\aB}\right) (\aH+1) - \aT - \aH' + \aH \right] + \aK \left[ 3 (\aM + \aH) - 2 \aB \left(1 - \frac{\aH'}{\aH}\right) \right. \nonumber\\
  & & \left. - \aT (2 \aB - \aH + 2) + \frac{\aH'}{\aH} - 2 \aH' + 2 \aM \aH + 1\right] - (\aH + 1) \aK', \\
 \epsilon_{5b} & = & 2 (\aB + 1)  \frac{\aH'}{\aH} - \aB (3 \aM + 2 \aT - \beta_3 + 5) + (\aH + 1) \left[ 3 (\aM + 1) - 2 \frac{\aB'}{\aB} - \beta_3 \right] \nonumber\\
  & & - 2 (\aT + 1), \\
 \epsilon_{5c} & = & 2  \left[\aM (\aH + 1) - \aT + \aH\right] \frac{\aB'}{\aB} \nonumber\\
  & & +(\aB-\aH)\left[2\left(\aM-\aT-\frac{H'}{H}\right) - (6\aB\aH+\aK)\frac{\rhom}{\alpha H^2M^2} \right]\frac{\aH'}{\aH} \nonumber\\
  & & + \aB \left[\aT - \aT' + (\aM + \aT + 3) (\aM - \beta_3) + \beta_3 - 2 (\aM - 2 \aT - 1)\frac{\aH'}{\aH} - \aM' + 2\right] \nonumber\\
  & & - \left\{ 2 (\aM - \aT)\frac{\aH'}{\aH} + 2 \aB' (\aT + 1) + \aH \left[\aM (\aM - \beta_3 + 2) - \aM' - \beta_3 + 1\right] \right. \nonumber\\
  & & \left. - \aM' + 2 \aH' (\aT + 1) + \aT' + \aT (\beta_3 - 1) + \aM (\aM - \aT - \beta_3 + 1) \vphantom{\frac{1}{1}} \right\}.
\eqa

\paragraph{Semi-dynamical coefficients}

We provide here the coefficients of the semi-dynamical modifications $\mu(a,k)$ and $\gamma(a,k)$, Eqs.~(\ref{eq:mu}) and (\ref{eq:gamma}), respectively, in Sec.~\ref{sec:mod}.
The numerator of $\mu(a,k)$ and denominator of $\gamma(a,k)$ are determined by
\bqa
 \mu_{+2} & = & \left\{-2 \aB \left[\bonfot + \fPhi (\bontwt + \bont \fPhi) + \bont \fPhi'\right] \gni \right. \nonumber\\
  & & \left.  + \alpha \left[-2 \aH \eBthr \left(\bonfot + \fPhi \{\bontwt + \bont \fPhi \} + \bont \fPhi'\right) + \bonsit (\gon + \aT \gon + \fPhi \gnit)\right]\right\}\frac{H'}{H} \nonumber\\
  & & -2 (\alpha \aH \eBthr + \aB \gni) \left[ \bont \fPhi \left(\frac{H'}{H}\right)^2 + (\bonsieit + \bonsit)(\fPhi - \fzeta)\frac{\rhom}{2H^2M^2}\right], \\
 \mu_{+4} & = & \left\{-2 \aB \left[\bonfit - \bonsit + (\aB - \aH)^2 (\fPhi \{\btht + \fPhi\} + \fPhi')\right] \gni \right. \nonumber\\
  & & + \alpha \left[\aB \aH \left(\bonsit \fPhi + 4 \aH \eBthr \left\{\fPhi [\btht + \fPhi] + \fPhi'\right\} - 2 \bset \left\{\gon + \aT \gon + \fPhi \gnit\right\}\right) \right. \nonumber\\
  & & + \aB^2 \left(\{1 + \aT\} \bonsit - 2 \aH \eBthr \left\{\fPhi [\btht + \fPhi] + \fPhi'\right\} + \bset \{\gon + \aT \gon + \fPhi \gnit\}\right) \nonumber\\
  & & + \aH \left(-2 \bonfit \eBthr + \bonsit \{\eBon - \aH \fPhi\} \right. \nonumber\\
  & & \left. \left. \left. + \aH \left\{-2 \aH \eBthr \left[\fPhi (\btht + \fPhi) + \fPhi'\right] + [1 + \aT] \bset \gon + \bset \fPhi \gnit\right\}\right)\right]\right\}\frac{H'}{H} \nonumber\\
  & & -2 (\aB - \aH)^2 (\alpha \aH \eBthr + \aB \gni) \left[ \fPhi \left(\frac{H'}{H}\right)^2 + (\bsenit + \bset)(\fPhi - \fzeta)\frac{\rhom}{2H^2M^2}\right],
\eqa
where we have defined $\mu_{+6} = \mu_{\infty}^+ + \aH \left( \fPhi\mu_{\Phi,\infty}^+ + \fzeta\mu_{\zeta,\infty}^+ \right)$ in Eqs.~(\ref{eq:muinfp}) through (\ref{eq:muzetainfp}).
The denominator of $\mu(a,k)$ is given by
\newpage
\bqa
 \mu_{-0} & = & \alpha \bont \left\{\bonfot + \bontwt \fPhi + \bont \left[\fPhi\left(\fPhi+\frac{H'}{H}\right) + \fPhi'\right]\right\}\frac{H'}{H} \nonumber\\
  & & + \left\{ 2 \aB \left[\bonfot + \fPhi (\bontwt + \bont \fPhi) + \bont \fPhi'\right] \gni \right. \nonumber\\
  & & + \alpha \left[ 2 \aH \eBthr \left(\bonfot + \fPhi \{\bontwt + \bont \fPhi\} + \bont \fPhi'\right) + \bonsieit (\{\bont + \gnit\} \fPhi + \gon\{\aT + 1\} \right. \nonumber\\
  & & \left. \left. - \bont \fzeta)\right] + 2 \bont \fPhi [\alpha \aH \eBthr + \aB \gni]\frac{H'}{H} \right\}\frac{\rhom}{2H^2M^2}, \\
 \mu_{-2} & = & \alpha \left\{\bonfit \bont + [\aB - \aH]^2 \left[\bonfot + \bontwt \fPhi + \bont \fPhi \left(\btht + 2 \fPhi + 2 \frac{H'}{H}\right) + 2 \bont \fPhi')\right]\right\}\frac{H'}{H} \nonumber\\
  & & + \left\{ 2 \aB \left[\bonfit + \bonsieit + (\aB - \aH)^2 \left(\fPhi \{\btht + \fPhi\} + \fPhi'\right)\right] \gni \right. \nonumber\\
  & & + \alpha \left[-\aB \aH \left(2 \left\{2 \aH \eBthr \left[\fPhi (\btht + \fPhi) + \fPhi'\right] + [1 + \aT] \bsenit \gon + \bsenit \fPhi \gnit \right. \right. \right. \nonumber\\
  & & \left. \left. + \bont \bsenit [\fPhi - \fzeta]\right\} + \bonsieit \{\fPhi - 2\fzeta\}\right) + \aB^2 \left(2 \aH \eBthr \left\{\fPhi [\btht + \fPhi] + \fPhi'\right\} \right. \nonumber\\
  & & \left. + \{1 + \aT\} \bsenit \gon + \bsenit \fPhi \gnit + \bont \bsenit \{\fPhi - \fzeta\} + \bonsieit \{1 + \aT + \fPhi - \fzeta\}\right) \nonumber\\
  & & + \aH \left(\bonsieit \eBon + 2 \bonfit \eBthr + \aH \left\{\bont \bsenit \fPhi + 2 \aH \eBthr \left[\fPhi (\btht + \fPhi) + \fPhi'\right] \right. \right. \nonumber\\
  & & \left. \left. \left. + \bsenit [\gon + \aT \gon + \fPhi \gnit]\right\} - \aH \{\bonsieit + \bont \bsenit\} \fzeta\right)\right] \nonumber\\
  & & \left. + 2 [\aB - \aH]^2 \fPhi [\alpha \aH \eBthr + \aB \gni] \frac{H'}{H} \right\} \frac{\rhom}{2H^2M^2}, \\
\mu_{-4} & = & \alpha (\aB - \aH)^2 \left\{ \bonfit + \bont \cs^2 + [\aB - \aH]^2 \left[\fPhi \left(\btht + \fPhi + \frac{H'}{H}\right) + \fPhi'\right] \right\} \frac{H'}{H} \nonumber\\
  & & + \left\{ 2 \aB [\aB - \aH]^2 [\bsenit + \cs^2] \gni + \alpha \left[\aB^2 \aH \left(\bsenit \eBon + 2 \cs^2 \eBthr \right. \right. \right. \nonumber\\
  & & \left. + \aH \bsenit \left\{1 + \aT + 3 \fPhi - 6 \fzeta\right\}\right) - \aB^3 \aH \bsenit (2 + 2 \aT + 3 \fPhi - 4 \fzeta) \nonumber\\
  & & + \aB^4 \bsenit (1 + \aT + \fPhi - \fzeta) + \aH \left(\aH^2 \left\{\bsenit \eBon + 2 \eBthr \cs^2 \right\} \right. \nonumber\\
  & & \left. + \en \{\bont \fPhi + \gon[1 + \aT] + \fPhi \gnit\} - \{\aH^3 \bsenit + \bont \en\} \fzeta\right) - \aB \aH^2 \left(4 \eBthr \cs^2 \right. \nonumber\\
  & & \left. \left. \left. + \bsenit \{2 \eBon + \aH[\fPhi - 4\fzeta]\}\right)\right]\right\}\frac{\rhom}{2H^2M^2},
\eqa
where we have defined $\mu_{-6} = \mu_{\infty}^- + \aH \left( \fPhi\mu_{\Phi,\infty}^- + \fzeta \mu_{\zeta,\infty}^- \right)$ in Eqs.~(\ref{eq:muinfm}) through (\ref{eq:muzetainfm}).
Finally, the numerator of $\gamma(a,k)$ in Eq.~(\ref{eq:gamma}) is given by
\bqa
 \gamma_{+0} & = & \alpha \bonsit \bont \frac{H'}{H} + 2 (\bonsieit + \bonsit) (\alpha \aH \eBthr + \aB \gni) \frac{\rhom}{2H^2M^2}, \\
 \gamma_{+2} & = & (\aB - \aH)^2 \left[\alpha (\bonsit + \bont \bset)\frac{H'}{H} + 2 (\bsenit + \bset) (\alpha \aH \eBthr + \aB \gni)\frac{\rhom}{2H^2M^2}\right],
\eqa
where we have defined $\gamma_{+4} = \gamma_{\infty}^+$ in Eq.~(\ref{eq:gammainfp}).


\vfill
\bibliographystyle{JHEP}
\bibliography{semidynamics}

\end{document}